\documentclass{aastex63}

\usepackage{fixltx2e}
\usepackage[shortlabels]{enumitem}
\usepackage[document]{ragged2e}
\usepackage[english]{babel}
\usepackage[autostyle]{csquotes}
\usepackage{amsmath}
\usepackage{bm}

\submitjournal{ApJ}

\shorttitle{Nova-produced Common Envelope}
\shortauthors{}
\graphicspath{{./}{figures/}}
\begin{document}

\title{Nova-produced Common Envelope: Source of the Non-solar Abundances and an Additional Frictional Angular Momentum Loss in Cataclysmic Variables}

\author{Warren M. Sparks}
\affil{210 La Mesa Lane, Georgetown, TX 78628; retired LANL}
\email{warrencharlene@suddenlink.net}
\author{Edward M. Sion}
\affil{Department of Astrophysics \& Planetary Science, Villanova University, \\
800 Lancaster Avenue, Villanova, PA 19085, USA}
\email{edward.sion@villanova.edu} 

\begin{abstract}

A substantial fraction of Cataclysmic Variables (CVs) reveals non-solar abundances. A comprehensive list of CVs which includes those that have been examined for these abundances is given. Three possible sources of these non-solar abundances on the secondary are accretion during the red giant common envelope phase, an Evolved Main Sequence secondary and nova-processed material. Use of the secondary's cross-section just on the escaping nova material to change the abundances of its convective region has been the killing objection for considering nova-processed material. The key element, ignored in other studies, is that a thermonuclear runaway on a white dwarf causes a strong propagating shock wave which not only ejects material, but also produces a large amount of non-ejected material which forms a common envelope. This nova-produced common envelope contains a large amount of non-solar material. We demonstrate that the secondary has the capacity and time to re-accrete enough of this material to acquire a significant non-solar convective region. This same envelope interacting with the binary will produce a Frictional Angular Momentum Loss which can be the Consequential Angular Momentum Loss needed for the average CV white dwarf mass, WD mass accretion rates, the period minimum, the orbital period distribution, and the space density of CVs problems. This interaction will decrease the orbital period which can cause the recently observed sudden period decreases across nova eruptions. A simple, rapid evolutionary model of the secondary that includes the swept-up nova-produced material and the increasing convective region is developed and applied to individual CVs.	

\end{abstract}


\section{Introduction} 

\setlength{\parindent}{4em}
When the more massive component of a long-period binary expands and engulfs its lower mass companion in a common envelope (CE) and its core and the companion spiral inward, a short-period binary can form \citep{pac76} inside the CE.  Eventually this CE is ejected leaving a pre-cataclysmic variable (pre-CV) consisting of a white dwarf (WD) primary and a red dwarf star secondary. The pre-CV loses angular momentum by gravitational radiation \citep{kol96} and magnetic braking of the secondary \citep{fau71,rap82} until the Roche lobe shrinks to where the secondary overflows and its material passes through the $1^{st}$ Lagrangian point (L1) and is eventually accreted onto the WD usually through a disk surrounding it.  Thus, a cataclysmic variable (CV) is born. In the case of a WD with a very strong magnetic field, a polar, this material is channeled directly onto its magnetic poles. If the ring material can accrete rapidly from a thermal viscous instability in the ring, a dwarf nova (DN) results and if this material accretes steadily, it is termed nova-like. The WD can have a core structure dominated by either He, CO or ONeMg. After enough hydrogen-rich material is accreted onto the WD, a thermonuclear runaway (TNR) will occur \citep{sta71}. The high temperatures reached causes the rates of proton capture and positron decay for $^{13}$N to be comparable, ie. the hot CN cycle. The outer layers of the WD mix with the accreted material either during the accretion process or during the TNR to form a He, CNO or ONeMg novae. The continuing loss of angular momentum from the CV by gravitational radiation and magnetic braking of the secondary causes its period to decrease. Magnetic braking causes heating in the secondary's outer layers driving them out of thermal equilibrium. The secondary's mass loss decreases the mass ratio ($q=M_{2}/M_{1}$) since the WD's mass doesn't change much. When the binary period decreases to $\sim 3 $ hr., the widely accepted viewpoint is that the convective region of the secondary extends all the way to its center disrupting its magnetic field \citep{dan82,rap83}. This disruption stops the magnetic braking so that the secondary's radius shrinks as the secondary comes into thermal equilibrium. When the secondary's radius falls below its Roche lobe radius, the secondary ceases to overflow its Roche lobe. The binary no longer appears as a CV, but it continues inexorably to lose angular momentum by gravitational radiation. When the secondary's Roche lobe shrinks to the volume of the secondary, overflow begins again at $\sim 2 $ hr. This striking paucity of CVs between orbital periods of 2 and 3 hours is known as the CV period gap. Gravitational radiation is normally assumed to be the dominant means of removing orbital angular momentum from the CVs below the period gap. After the secondary loses its hydrogen-rich envelope, the period begins to increase, which is known as the period bounce \citep{rap82}. 

The composition of the material observed from CVs before the bounce is expected to be, and normally is, solar. Yet, roughly 15\% of CVs observed in the UV \citep{gan03} reveal abnormal (non-solar) abundances. Usually, these abundances indicate that the material has undergone thermonuclear (TN) processing by the CN cycle.
These observations include:

\begin{enumerate}[(a)] 
\item \textbf{$^{12}$CO/$^{13}$CO}. This is perhaps the most direct indicator of TN processing because it is independent of the carbon abundance. It is generally agreed that the $^{12}$CO/$^{13}$CO (actually, the $^{12}$C/$^{13}$C) ratio is due to the CN cycle. Scott et al. (2006) gives a solar $^{12}$C/$^{13}$C value of 86.8, while if the slow CN cycle is run to completion this ratio is 4 (Wollman 1973).  
\item \textbf{NV 1240/CIV 1541}.  These far ultraviolet emission resonance lines are similar in atomic structure and should directly give the N/C ratio. The solar value is 0.25 (Grevesse et al. 2011), while the equilibrium for the slow CN cycle approaches 94 (Caughlan 1965).
\item \textbf{CO features in the infrared}. The reduction or absence of the CO features compared to an equivalent single star is assumed to be a carbon deficiency. 
\item \textbf{Far ultraviolet spectra following a DN outburst}. Although the metallic atoms of the accreted material onto the WD after a DN outburst will gravitationally diffuse downward, the relative abundance can be determined. Large overabundances of heavy metals are considered abundance anomalies. 
\item \textbf{Near-infrared lines}. Synthetic spectra are used to determine [Fe/H] and other metals relative to Fe on the secondary. Those with an underabundance of carbon relative to iron are considered non-solar, as are underabundances of magnesium and overabundances of sodium. 
\item \textbf{Infrared continuum}. In order to match the infrared continuum, significant hydrogen deficits are required which indicates recent H-burning. 
\end{enumerate}

In addition to the non-solar abundances discussed above, current evolutionary 
studies of CVs have yielded new information on the following topics: {\bf (1) the period minimum}, which occurs when the secondary becomes degenerate, then further mass loss causes its radius to expand and the orbital period increases, i.e. the period bounce, \citep{rap82}; {\bf (2) the orbital period distribution of CVs} above and below the period gap and the fraction of period bouncers; {\bf (3) the average mass of a white dwarf in a CV and the fraction of helium-core WDs}; {\bf (4) a robust, more accurate space density of CVs} from the SDSS data releases; and {\bf (5) the variation of the accretion rates onto the WD}. The recent conclusions of \citet{sch20} deliver the unexpected results that the period change across a classical nova outburst decreases instead of increasing.

 The CVs with mass ratios and/or non-solar abundances are collected and tabulated in Section 2. Section 3 examines the proposed sources of these non-solar abundances and their short-falls. The discrepancies between evolutionary studies and observations are also discussed here. Material that expands but not ejected from a nova outburst is explored as a source of the non-solar abundances and the Consequential Angular Momentum Loss proposed in many evolutionary studies in Section 4. A simple, rapid method of calculating the non-solar abundance build-up on the secondary is presented and compared with observations of individual CVs in Section 5. Section 6 finishes with the conclusions and discussion.

\section{The Observational Evidence for CVs with Non-Solar Abundances} 

The radiation from a CV may come from many different locations:  the donor secondary, the ballistic jet of material overflowing through the L1 point, the splash of the jet hitting the disk (hot spot), continuum radiation from wind outflow, the WD and the jet directly striking the WD in the case of a polar.  In general, infrared radiation is dominated by the secondary and ultraviolet radiation by the WD, hot spot and/or the jet. The CO features only appear in the infrared of late-type secondaries. If these features from spectra of the CV's companion stars are weaker or absent when compared to their equivalent single stars, it is normally assumed that the carbon abundance is below solar. The weakness or absence of the CO features is a fast way to identify a CV with non-solar abundances. \citet{ham11} has examined the infrared spectrum for the CO absorption features in many pre-CVs and CVs and combined them with other studies. Harrison and colleagues \citep{har04,har05,har17} have determined and listed the $^{12}$CO/$^{13}$CO ratio from the K-band spectra.  

Observations of the NV/CIV line flux from the IUE \citep{mau97} and from the Space Telescope Imaging Spectrograph are summarized by \citet{gan03,gan04}. The results for V2301 Oph and V1432 Aql from the Space Telescope \citep{sch01} are also included. 
Observations of the ultraviolet lines from the WD after a DN have not been summarized in the literature. Individually they are GW Lib \citep{Sz02a},
BW Scl \citep{Ga05}, LL And \citep{Ho02}, WZ Sge \citep{Si95a}, AL Com \citep{Sz03}, SW UMa \citep{Ga05}, HV Vir \citep{Sz02c}, WX Cet \citep{Si03},
EG Cnc \citep{Sz02c}, BC UMa \citep{Ga05}, VY Aqr \citep{Si03}, EK Tra \citep{Ga01,Go08}, VW Hyi \citep{Si95b}, EF Peg \citep{Ho02}, MV Lyr \citep{Ho04},
DW UMa \citep{Kn00}, WW Cet \citep{Go12}, U Gem \citep{Lo06,Go17}, SS Aur \citep{Si04}, RX And \citep{Si01}, RU Peg \citep{Si04,Go08}, BY Cam \citep{Mo03}, BV Cen \citep{Go12}, CH UMa \citep{Go12}, EY Cyg \citep{Go08}, EM Cyg \citep{Go12}, SS Cyg \citep{Go12}, UU Aql \citep{Go12}, SS Aur \citep{Go21}, TU Men \citep{Go21}, 
and BZ UMa \citep{Go11}.  Infrared synthetic spectra of K-type secondaries have been compared to 41 CVs in a massive study by \citet{har16,har17}.
\citet{har18} has also generated synthetic spectra with H-deficient stellar atmospheres and finds four systems that have significant hydrogen deficits. 
\footnote{
Other CVs, notably AM CVn binaries, have little or no hydrogen. \citet{nel10} give a list of these binaries with the composition of their secondaries and discuss three different scenarios for their formation. For our purposes we have not included AM CVns in our list of non-solar abundant CVs based solely on having little or no hydrogen.
}

Table 1 contains all the CVs, whose period ($P$) and mass ratio ($q$) are known from the \citet{rit03} catalogue (version7.24, updated to 31 Dec. 2015 and additions listed in its footnote) and all of the CVs examined for abundance anomalies.  We did not include CVs whose \textit{q} value is estimated or an upper or a lower limit. The different anomalies are shown in six different columns. A \enquote{Y} indicates an abundance anomaly was observed while a \enquote{N} means it was looked for but not found. Thus, our table designations for the CO absorption features are opposite from \citet{ham11} because a \enquote{N} in their table 4 indicates no CO features were found. \enquote{V} stands for a variable feature. A \enquote{YH} designates enhanced abundances of odd-numbered nuclear species above oxygen indicating proton capture on even-numbered nuclear species during high temperature CNO or ONeMg H-burning. The carbon abundances determined from infrared spectra synthesis that give [C/FE] $\le$ 0.0 are listed as a \enquote{N}. Data that is questionable because of low S/N or emission in absorption features are not included. First of all, this table shows that there are a lot of CVs that need further study.  The table reveals that the abundance anomalies appear more often in the longer period CVs than in the shorter period ones. This trend has been noted by \citet{har16} for the infrared synthetic spectra studies. It is not known if the CVs with abnormal abundances have longer Ps and higher qs or if this is a selection effect. Tappert et al. (2007) has shown that pre-CVs do not have abnormal CO bands or large $^{13}$CO abundances.  \citet{how10} has noted that pre-CVs and polars (or AM Her stars) appear to have normal CO absorption bands and after examining the K-band spectra of five polars, \citet{har05b} title their paper \enquote{Why are the Secondary Stars in Polars so Normal?}. However, two of the five magnetic CV found \citet{gan03} are polars with enhanced NV/CIV ratios.

\section{The Problems}

We have summarized ample observational evidence that the secondary stars of a significant fraction of CVs contain strikingly non-solar metal abundances associated with thermonuclear processing. Under the current theory of CV evolution, the secondary is not expected to possess this material. There are three possible sources of CN cycle material on the secondary:  {\bf 1)} accretion during the common envelope (CE) following the expansion of the more massive giant primary which engulfs the secondary.  The secondary spirals in, strips the primary's outer envelope and accretes some nuclear-processed material, while forming a close binary. {\bf 2)} the secondary already has nuclear-processed material from its Evolved Main Sequence (EMS). The CV originally had a donor previously more massive than the primary leading to a short phase of unstable thermal timescale mass transfer (TTMT). The secondary has its outer layers stripped during the TTMT revealing core-nuclear-processed material. {\bf 3)} accretion of nuclear-processed material from a nova explosion on the WD. This can occur from the intercept of the material ejected during the nova eruption and from the nova-produced common envelope (NCE). The latter is not to be confused with the red giant common envelope (CE) event described earlier. 

Models \citep{Hj91} of the CE event show that only a small net amount of material (typically $\le0.01 M_{\odot}$) is accreted and retained by
a 1.25$M_{\odot}$ secondary.  To cover a wider mass range of accretion, \citet{mar98} assumed that $0.2M_{\odot}$ was accumulated and presented an expansive theoretical study of enhancements on massive secondaries due to the CE stage and nova eruption.  Initially, material accreted during the CE phase and nova outbursts is mixed inward by convection and thermohaline mixing, and it dominates the secondary's surface layers. Gradually, material accreted from the CE is diluted during secondary overflow and the novae ejecta dominates. As the surface convection penetrates to deeper layers, the secondary's surface composition is dominated by the original secondary's composition until the secondary has lost most of its mass. If only the accretion from the CE is considered, then any abundance enhancement is rapidly diminished. Only the geometrical cross-section of the secondary was considered for the capture of the nova ejecta.  \citet{ste99} studied the chemical evolution of population I and population II secondaries due to the capture of nova ejecta. They find that the metallicity of the secondary increases in population I secondaries only if the accreted amount is an order of magnitude larger than by its geometrical cross-section. For population II, the nova ejecta dominates even if the accumulation is an order of magnitude smaller than the geometrical cross-section. The EMS scenario \citep{pod03} allows for CVs in the period gap and below the period minimum and a large range of secondary masses with periods above 5hr. Binary Population Synthesis (BPS) studies \citep{how01a,how01b} found that it is unlikely that He burning prior to the secondary reaching its Roche lobe can be an important component leading to non-solar metal abundances in the secondary. However, these BPS models do not include an angular momentum loss mechanism. The approach of \citet{kal16} was to assume a large grid (initial $\bm{M_{1}}$, initial $\bm{M_{2}}$, orbital period) of all possible configurations in order to not exclude any possible paths using the binary stellar evolution code MESA. These paths do include the EMS scenario. Both the CE phase and the EMS scenario do not explain the pre-CVs or polars below the period gap not having abundance anomalies. The upper limit for the donor star of $4.7M_{\odot}$ determined by \citet{kal16} for the EMS scenario is much lower than the initial mass of $>8M_{\odot}$ \citep{ibe84} required for abundances enhancements of elements above Oxygen. These enhancements are observed in the ultraviolet spectra \citep{Lo06,Si95b} and in the infrared spectra \citep{har16} for some of non-solar CVs designated in table 1 with \enquote{YH}. 

The difficulties with the secondary capturing enough nova-processed material via a geometrical cross-section to dominate its surface layers are:

\begin{enumerate}
    \item Its geometrical cross-section is small ($<3$\%). The difficulty of not capturing enough nova ejected material is emphasized by calculations of \citet{ste99} which required $10 \times$ the geometrical cross-section to increase the metal abundances of the secondary's surface layers.
    \item Thermohaline mixing and convection mixes the re-accreted material inward \citep{mar98}, thus diluting the nuclear-processed material.
    \item The high-velocity material that the secondary geometrically intercepts is not highly nuclear-processed.
    \item The non-solar abundance of this material is also diluted by the accretion disk around the WD \citep{fig18}. 
\end{enumerate}

The average WD mass in CVs and pre-CVs should be smaller than the average WD mass ($ \le0.6 \ M_{\odot}$)  \citep{pol96} of single stars because the presence of the secondary helps eject the CE which stops the mass growth of the WD. There should also be a larger fraction of helium-core WDs  \citep{pol96,how01a} than for single WDs. However, the CV's average WD mass is observed to be $0.83 \ M_{\odot}$ \citep{mca19,pal20}. They also do not observe any He-core WDs in CVs.  This discrepancy cannot be solved by a WD mass growth from the nova outburst cycle, the CE phase or EMS scenarios \citep{wij15,sch16}. The WD mass being virtually the same above and below the period gap \citep{mca19} is also a strong argument against an increasing WD mass during the CV phase. The pre-CVs have WD masses and fractions of helium-cores similar to single WDs \citep{zor11}. {\bf These observations lead to the argument that CVs with lower mass WD must become dynamically unstable} \citep{nel16,zor19}. \citet{sch16} find that BPS models with consequential angular momentum loss (CAML) that increases with decreasing WD mass give agreement with observations. They also find that this CAML brings better agreement with the space density and orbital period distribution of CVs and speculate that Frictional Angular Momentum Loss (FAML) following a nova outburst could cause this CAML. Nelemans et al. (2016) assumed that the interaction between the expanding nova envelope and the secondary leads to a common-envelope-like phase that could take away angular momentum from the binary. They calculated that low-mass WDs must eject $\sim$40\% of the nova material by this mechanism for the mass transfer to be unstable. \citet{liu16} calculated that if 20-30\% of the material ejected during a nova eruption remains in the system in the form of a circumbinary disk, CVs with low-mass WD would more likely become dynamically unstable.

BPS studies \citep{how01a,gol15} predict that a large majority of CV systems should be 
below the period gap with $\sim1/2$ being period bouncers. However, \citet{pal20} 
found only 7\% to be period bouncers in a volume limited sample of CVs. 
The predicted space density of CVs from BPS evolutionary models \citep{kol93} is 1-2 orders of magnitude larger than observed \citep{pal20}.  
After the CV evolves through the period gap, gravitational radiation was thought to be the only mechanism to remove orbital period angular momentum. Using gravitational radiation as the only angular momentum loss below the gap, \citet{gol15,kal16} predicted a minimum orbital period of 65-70 min while the observed value is 76-82 min \citep{kni11,mca19}. \citet{kni11} found that gravitational radiation must be scaled up to $2.47\pm.02$ for the angular momentum loss below the gap to produce period distributions and a period minimum that 
agrees with observations. Pala et al. (2017) investigated the effective 
temperatures of WDs in CVs and obtained a good fit to observations when there is an additional angular momentum loss below the gap. However, \citet{liu19} calculated the FAML on the secondary from nova expanding envelopes for several constant velocities and demonstrated that FAML was unlikely to account for the increase of angular momentum loss above gravitational radiation unless the expanding velocities are extremely low.  They did not include the FAML between the binary and the NCE. As the CVs evolve, simple theory predicts their evolutionary paths in the period and secondary's Roche lobe overflow rate space should converge \citep{rap83,kol01}. Since the WD's mass accretion rates are equal to or a fixed fraction of the overflow rate, they should also converge. However, the spread of the observed WD temperatures at a given period indicates that their WD's accretion rate varies more than theory predicts. Theory also predicts that the secondary's Roche lobe overflow rates will decrease as the binary period decreases. Thus, one would expect a separation at some period between the nova-likes, which accrete onto the WD at a high rate, and the DNs \citep{kni11}. In contrast, the nova-likes and DNs are observed to overlap over a long period range.

The increase in binary separation and orbital period from a 
nova outburst has long been predicted \citep{sha86} and accepted. Basically, it comes from the loss of mass and angular momentum from the WD primary in the CV system. This period increase dominates over the decreases due to capture of ejected material by the secondary and FAML \citep{mac86}, when the ejected material flows by the secondary.  In fact, the hibernation theory \citep{sha86} is built on the premise of a period increase from a nova outburst. \citet{sch20}, using accurate pre- and post-nova periods, has shown that the period actually decreases for five of six classical novae.  These observations demand another 
source of period decrease by the CV system during a nova outburst. 
   
\section{The Solution}

The rapid increase of energy during a TNR causes a strong shock wave to develop in the outer layers of the WD \citep{sta74} in the CV. As it propagates through the decreasing density layers, it leaves behind a {\it large velocity gradient} in the expanding envelope (Sparks 1969). Part of this envelope can reach escape velocity and the remainder is gravitationally bound to the CV system. Part of this remaining shell will be gravitational bound to just the WD, and part will become a nova-produced common envelope (NCE) (see also Livio et al. 1990) that is not initially co-rotating with the CV. Starrfield, Truran \& Sparks (1978) predicted nuclear burning time scales of $\sim$100 yr. for the material on the WD, but the observed time scale is $\sim$10 yr (MacDonald, Fujimoto, \& Truran 1985).  

We will now explore the possibility that nova-processed material collected during the NCE phase is a source for the non-solar abundance CVs. But first we must be sure that there is enough capacity or room on the secondary for this material, i.e. can the secondary absorb this material and still remain within its Roche lobe? \citet{sha86} has shown that the ejection of material from the WD during a nova causes the binary separation to increase and, consequently, the secondary's Roche lobe to expand. Their equation (14) is 

\begin{equation}
\frac{\Delta a}{a} = \frac{\Delta M_1}{M_ 1} 
 \left[
\frac{(1 + 2 \beta q - \beta)}{ (1+q)} - \frac{2 \beta }{ q(q+1)}
\right],          
\end{equation}

where $a$ is the binary separation, $M_{\rm 1}$ is the mass of the WD, $\Delta M_{\rm 1}$  is the mass ejected from the binary system, $\beta$ is the fraction of the ejected nova mass collected on the secondary by the geometrical cross-section. 	For the large ejection velocities found in novae and assuming spherically symmetric ejection, they find that the geometric cross-section ($\beta$) of the equivalent Roche lobe radius is usually of order 10$^{-2}$, and $\beta\le0.036$ for $q\le1$. 
We will now calculate the maximum amount or capacity of nova-ejected material that can eventually end up on the secondary, which we call $\gamma$. Thus, we are including {\it any} nova-ejected material that is captured by the secondary. Since $\beta$ and $\gamma$ are both the fraction of ejected material on the secondary, the total angular momentum gives the same equation as eq. 1, i.e.,   

\begin{equation}
\frac{\Delta a}{a} = \frac{\Delta M_1}{ M_1} 
 \left[
\frac{(1 + 2 \gamma q - \gamma)}{(1+q)} - \frac{2 \gamma }{ q(q+1)}
\right].
\end{equation}

\citet{pat05} finds an empirical mass-radius relationship of

\begin{subequations}
\begin{align}
    R_2 =0.92 M_2^{0.71} \\
    \intertext{above the period gap and}
    R_2 =0.62 M_2^{0.61}
\end{align}
\end{subequations}

\noindent
below the gap. Therefore, 

\begin{subequations}
\begin{align}
    \frac{dR_2}{R_2} = 0.71 \frac {dM_2}{M_2}\\
    \intertext{above the gap and}
    \frac{dR_2}{R_2} = 0.61 \frac{dM_2}{M_2}
\end{align}
\end{subequations}

\noindent
below the gap. Using $q= \frac {M_2} {M_1}$, and $dM_2=+ \gamma \Delta M_1$, where $\Delta M_1$ is positive,

\begin{subequations}
\begin{align}
    \frac {\Delta R_2} {R_2} = 0.71 \frac {\gamma}{q} \frac{\Delta M_1}{M_ 1}\\
    \intertext{above the gap and}
    \frac {\Delta R_2} {R_2} = 0.61 \frac {\gamma}{q} \frac{\Delta M_1}{M_ 1}
\end{align}
\end{subequations}

below the gap. Taking Iben \& Tutukov's (1984) expression for the equivalent Roche lobe radius for the secondary,  

\begin{equation}
R_{\rm L2} = 0.52a \left[\frac {M_2} {M_1 + M_2} \right]^{0.44}.
\end{equation}
We find     

\begin{equation}
\frac{\Delta R_{\rm L2}} {R_{\rm L2}} 
= 
\frac{\Delta a}{a} + 0.44 \frac{ \Delta M_1}{M_1}
\left[ 
\frac{\gamma} {q} + \frac{1}{(1+q)} - \frac{\gamma}{(1+q)} 
\right].   
\end{equation}

Using the equivalent Roche lobe radius as an upper limit for the radius of the secondary, combining 
eqs. (2), (5a) or (5b), and (7), and solving for $\gamma$, we derive 

\begin{subequations}
\begin{align}
    \gamma = \frac{1.44q}{0.27(1+q) -2 q^2 + 1.44q + 2}\\
    \intertext{above the gap and}
    \gamma = \frac{1.44q}{0.17(1+q) -2 q^2 + 1.44q + 2}
\end{align}
\end{subequations}

\noindent
below the gap.

For $q = 1.0$ and above the gap, $\gamma$ = 0.73, and for $q = 0.1$ and below the gap, $\gamma$ = 0.067. The weak point of this derivation is in the use of the empirical mass-radius relationship for secondaries. Although this relationship probably applies over the long term evolution of the secondary, during the short time scale of re-accretion, the secondary probably is more like adiabatic. When mass is removed adiabatically from a star with a convective envelope, the radius can expand \citep{hje87,web88}. Thus, adding material adiabatically can make the radius shrink. Assuming an adiabatic behavior will change eqs 5a and 5b to

\begin{equation}
\frac {\Delta R_2} {R_2} = -0.33 \frac {\gamma}{q} \frac{\Delta M_1}{M_ 1} \tag{5c}
\end{equation}

and eqs 8a and 8b to

\begin{equation}
\gamma = \frac{1.44q}{1.23 + 0.67q -2 q^2} \tag{8c}
\end{equation}

For $q=1.0$, $\gamma = -13.9$, which means unlimited capacity, and for $q=0.1$, $\gamma=0.11$. From this analysis, we know that for larger $q$s, there is plenty of room or capacity for the secondary 
to assimilate nova-processed material, while at lower $q$s, the amount of room is starting to be a factor.

Next, we will examine the nova outburst to understand when (or opportunity) and how (or means) the secondary can accrete its nova-processed material. We will use a WD mass of $1.0M_{\odot}$  and a secondary mass of $0.6M_{\odot}$, which suggests a period of 3.43 hr. from empirical fits to CVs \citep{pat05}.  As the CV's angular momentum decreases due to magnetic braking and/or gravitational radiation, the Roche lobe of the secondary shrinks. This causes material from the secondary to flow through the L1 point (Fig.1a). This material can flow directly onto the WD if its magnetic field is strong enough (polars) or form an accretion disk around the WD (as shown in Fig.1a). In both cases hydrogen-rich material from the secondary is accumulated onto the WD until a TNR occurs [see \citet{sta71}] to give rise to a nova outburst, causing material to expand rapidly from the WD. Fig.1b shows how the geometrical cross-section of the secondary intercepts some of this material. For demonstration purposes, this figure represents the first nova outburst when the accreted material was solar. The expanding shell (blue) from the nova strikes the secondary (red) and mixes into the  outer convective region (light purple) of the secondary. The outer layers of the expanding material have a much higher velocity than the orbital velocity of the secondary (in this example 297 km/s). This geometrical cross-section capture by the secondary is the only re-accretion considered in most studies [for example, \citet{sha86}].

\begin{figure}
\vspace{-3.cm} 
\gridline{ 
          \fig{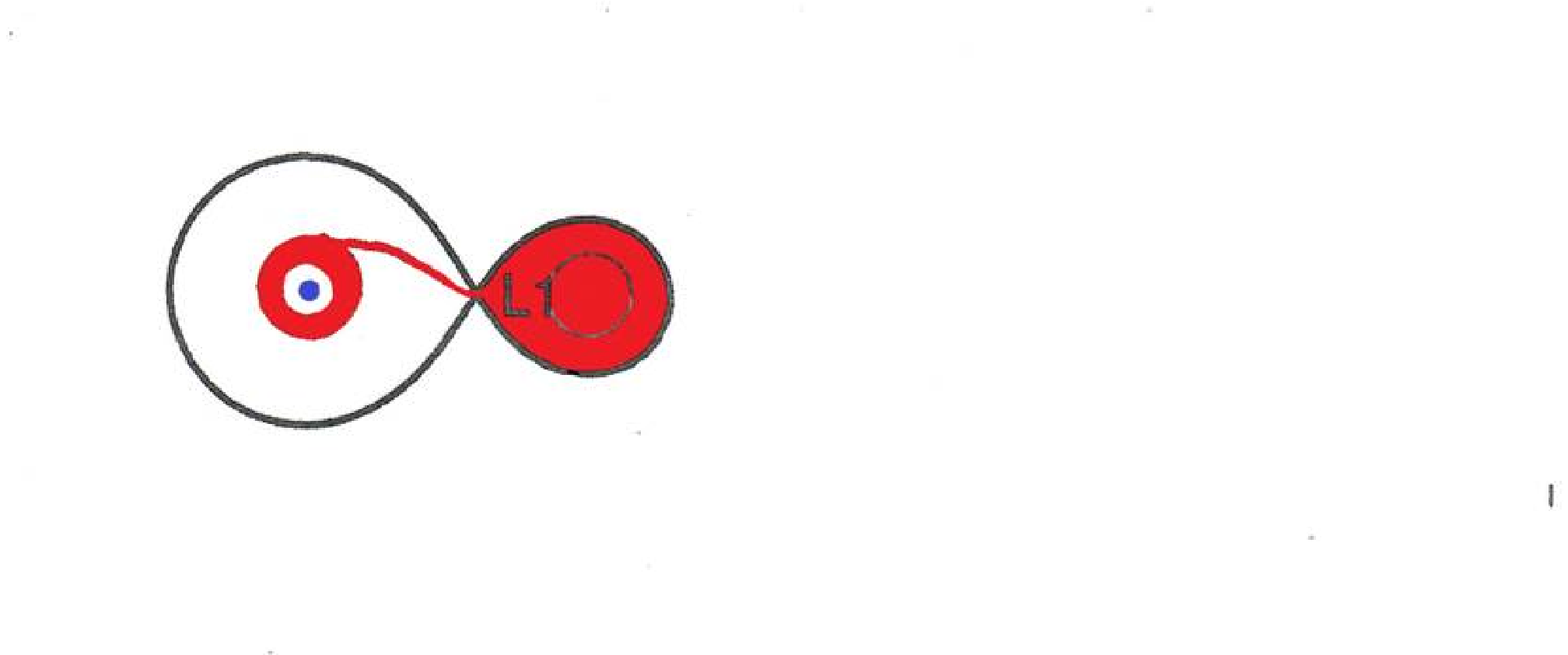}{0.5\textwidth}{1a} 
          \fig{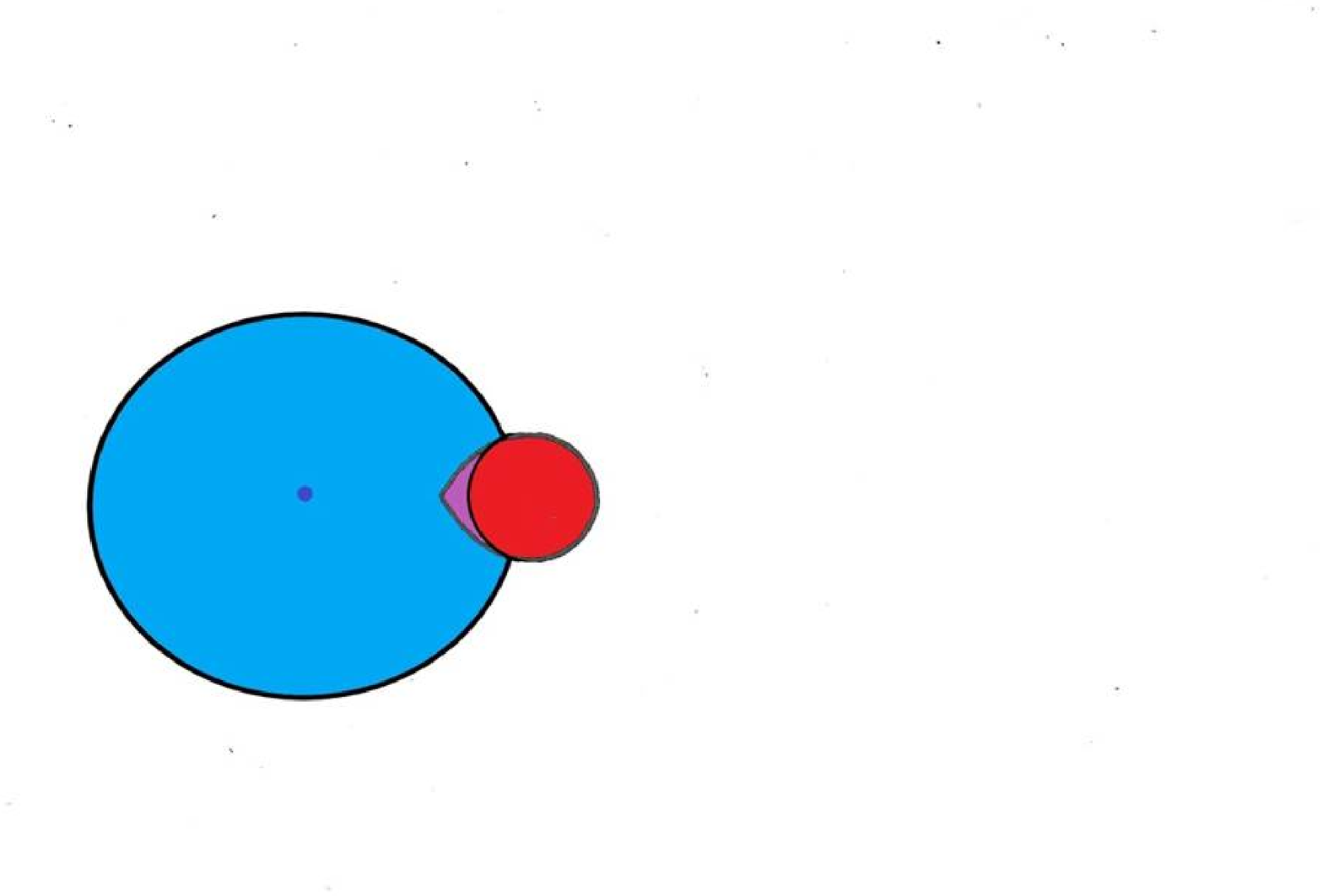}{0.5\textwidth}{1b} 
          }  
\vspace{1.cm} 
\gridline{ 
          \fig{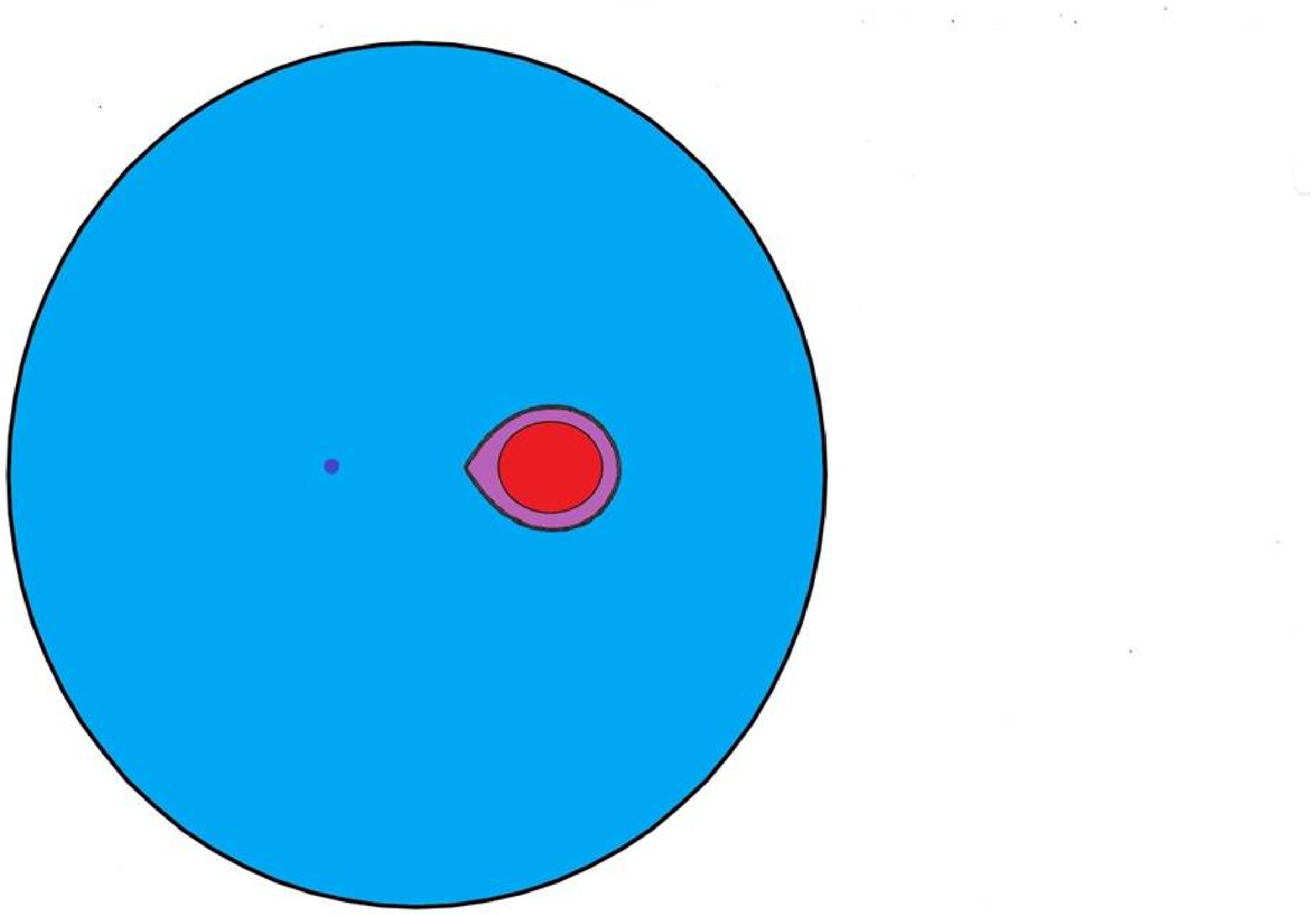}{0.5\textwidth}{1c} 
          \fig{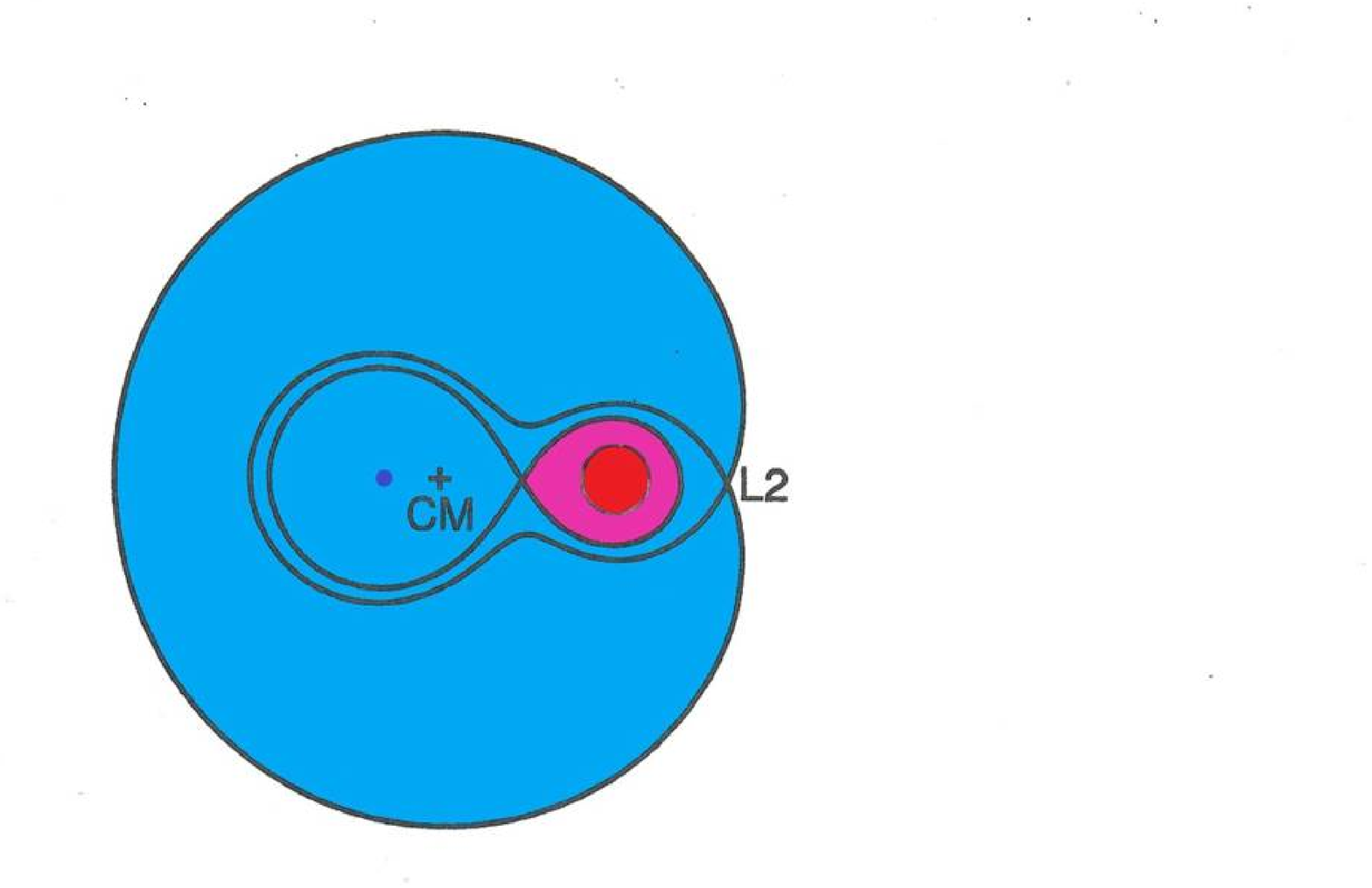}{0.5\textwidth}{1d} 
          }  
\vspace{-2.cm} 
\gridline{ 
          \fig{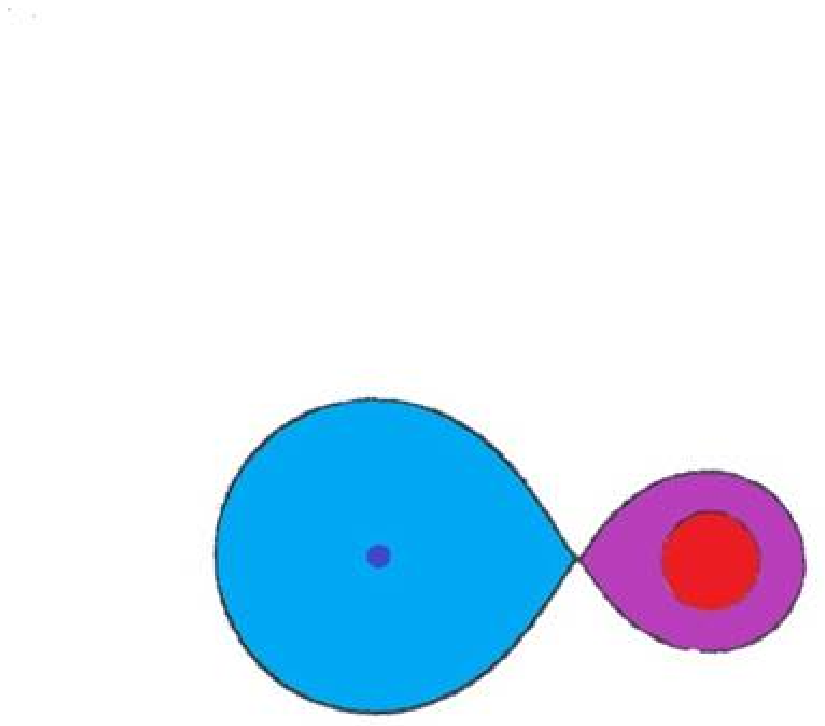}{0.5\textwidth}{1e} 
          \fig{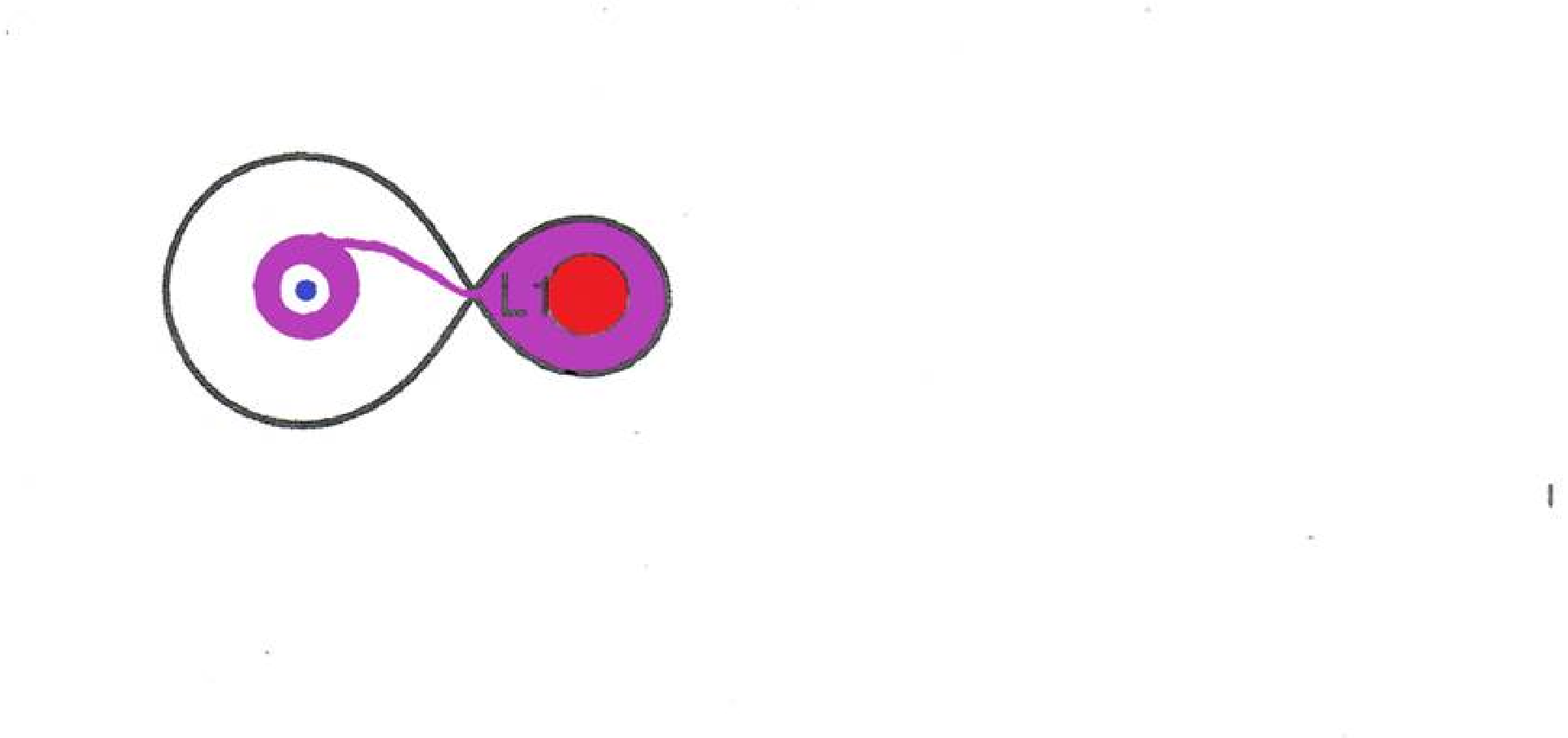}{0.5\textwidth}{1f} 
          }  
\caption{ Cross-sectional views of a CV. (a) The secondary (red) is overflowing its Roche lobe at the L1 point which forms a ballistic jet (red) that strikes the accretion disk (red) around the WD (dark blue).; 
(b) The WD (dark blue) undergoes a TNR and ejects nova-processed material (light blue). The secondary (red) intercepts and mixes this material inward by convection and thermohaline mixing into its convective region (light purple).;
(c) The non-rotating non-ejected NCE (light blue) surrounds the WD (dark blue) and the secondary (red) with its mixed convective envelope (light purple).; 
(d) The NCE (light blue), which now co-rotates with the WD (dark blue) and the secondary (red) with its mixed convective envelope (purple), has shrunk within the equipotential surface that contains the L2 point.; 
(e) The NCE has shrunk to just the Roche lobe (light blue) around the WD (dark blue). The secondary (red) with its mixed convective envelope (dark purple) is within its Roche lobe.; 
(f) The remaining NCE around the WD (dark blue) has been burnt and the secondary (red) is now overflowing its Roche lobe again and the nova-processed material from its mixed convective layer (dark purple) forms an accretion disk (dark purple) around the WD (dark blue).
}
\end{figure}

Hoyle \& Lyttleton (1939) studied the steady flow of material around a star.  The gravity of the star focuses the material into a wake behind it. The radius at which this material is accreted onto the star is the gravitational (Bondi-Hoyle) radius. Using the effective velocity of the secondary through the nova ejecta, the gravitational radius is normally smaller than its Roche Lobe radius. This Roche lobe's cross-section to a rapidly expanding shell is normally assumed for the secondary's re-accretion. \citet{sha86} studied the capture of nova-processed material with a constant expansion velocity included the geometrical cross-section of the secondary and the FAML interaction of the ejected material with the secondary. They found that the amount of captured material was small and that with FAML interaction, the binary separation increased except for very-low-q CV systems. 
 
 When the energy generation in the accreted layer of a WD exceeds $\sim10^{16}$ erg/gram/second, hydrodynamic nova models \citep{sta74,pri92,jos98} show that a shock wave will develop in it, propagate outward and eject material, leaving behind a large, hot NCE. \citet{pri86} describes such a model where convection lasts for 4.7 hr., which she showed is plenty long to thoroughly spread the nuclear-processed material throughout the NCE. We will use model 7 from \citet{sta74} because it is a published model where the structure of the NCE is plotted after almost all of the ejected material has escaped ($\sim$ 1 day). For this model when the expanding velocity is 570 km/s, the gravitational radius is equal to the Roche Lobe radius. After this time, the gravitational cross-section will dominate over the geometrical cross-section. The separation point between the escaping material and the non-escaping material due to just the TNR and energy from the $\beta^{+}$ unstable isotopes decay does not extend very far beyond the binary separation. Thus, at least for this example, the gravitational cross-section probably does not significantly increase the collection of nuclear-processed material on the secondary beyond the geometrical cross-section. However, a large amount of nuclear-processed material ($\sim 3.0\times 10^{29}$ g. or equal to 89\% of the ejected mass) is between the ejected material and the material gravitationally bound to just the WD, i.e., the NCE. At the base of this material the hot burning temperature drives convection to rapidly mix it. The C is converted to N via the hot CN cycle. Unlike the ejected material, the NCE's material continues to approach the equilibrium abundances which are determined by the temperature and density of the burning region. \citet{cau62} show that at the high temperatures of a TNR, the equilibrium N/C ratio can exceed 1000. As the temperature decreases, this ratio approaches $\sim$ 4 for the densities in the burning region. At still lower temperatures ($T_{6}\sim$ 50), this ratio increases and approaches the equilibrium N/C ratio ($\sim$100, \citet{cau62}) for the cold CN cycle.

Fig.1c shows the secondary (red) and its convective region (light purple) within the NCE (light blue). The material re-accreted is now mixed throughout the convective region. Using the density ($\bm{\sim 10^{-6}} \text{g/cm}^{3}$) of this NCE at the distance of the secondary, the secondary's orbital velocity (297 km/sec), which now dominates over any expansion or contraction of the NCE, and the geometrical cross-section of the secondary, the time for the secondary to rotationally sweep up $\bm{1.0 \times 10^{29}}$ g, which is $\sim$ 30\% of the ejected mass, is $\bm{\sim 12.5}$ d. This mass amount is what \citet{ste99} found that was needed for population I secondaries to increase their surface metal abundances. This time is probably a lower limit because the NCE must continually fill the space being swept up by the secondary. If we use the WD's Roche lobe equivalent radius as the upper boundary for its gravitationally bound material, then the average density is ($\bm{\sim 10^{-7}} \text{g/cm}^{3}$) in the NCE and we find the time for the secondary to rotational sweep up this amount is $\sim$ 1/3 yr. The actual time is probably between these two limits. The lifetimes of the WD's remnant burning (1-10 yr.) from observations (Krautter et al. 1996, Balman and Krautter 2001, Page et al. 2015, Vanlandingham et al. 2001) indicate that the NCE lasts long enough for the secondary to sweep up this amount of material. Just considering this amount of mass re-accreted by the secondary is not enough to give a negative period change according to eq(1). The models of \citet{pri86,pri92} on $1.25M_{\odot}$ WDs also produce high-velocity ejected material and less massive but significant envelopes larger than the typical binary separation ($\sim$ 10$^{11}$ cm). Their models have a mass remnant of 10 to 28\% of the ejected mass and a slowly expanding nebula (18 to 41\% of the total ejected mass) moving at a velocity of 30-70 km/s., which could also be partially captured. Although the mass amount is model and probably code dependent, it shows that the mass available for the secondary to accrete from the NCE is much more than just its geometrical cross-section of the high-velocity material. In summary, the NCE extends farther than the typical binary separation, is nuclear burning at its base, and is strongly convective. These properties allow this material to approach equilibrium burning much more (and thus, be more non-solar) than the high-velocity ejected material did. 

Thus, the NCE can potentially satisfy the four objections raised earlier against nova-processed material as the source of the non-solar abundance CVs. When material first leaves the WD, it is rotating around the binary's center of mass like the WD, but as it expands farther away from its surface it is not. Part of the NCE is blown away by winds, part is now ejected by FAML \citep{mac85}, and part is swept up by both components. As the CV binary rotates within the NCE, material striking the secondary will rapidly be mixed inward while material falling into the WD's Roche lobe is limited by the Eddington's luminosity. The CV will lose angular momentum as FAML ejects some material and, possibly from the secondary's magnetic field \citep{mar11}, causing some of the NCE to co-rotation with it. Forcing the NCE to co-rotate with the binary system takes angular momentum from the binary orbit and the orbital period decreases. This binary angular momentum loss may be the CAML needed for some BPS evolutionary studies to agree with observations.

Using the assumption that angular momentum lost from the WD during a nova outburst is symmetrical about its rotational axis \citep{sha86}, the orbital angular momentum lost is equal to $M_{ej}v_{1}r_{1}$, where $M_{ej}$ is the mass ejected, $v_{1}$ is the orbital velocity of the WD around the center of mass (= $2\pi r_{1}/P$), \textit{P} is the orbital period, and $r_{1}$ is its distance from the center of mass. This mass loss causes the period to increase \citep{sha86}. For our example of a $1.0M_{\odot}$ WD and a $0.6M_{\odot}$ secondary, $r_{1}$ is 0.375a, where a is the binary separation. We will use the angular momentum at the distance of the 2nd Lagrangian point (L2) from the center of mass [$r_{L2}$ = 1.613a from \citet{moc84}] as representative of the angular momentum given to the material in the NCE (see fig.1d) since material at this point has received enough angular momentum to escape. If the rotating NCE's mass exceeds 5.4\% of $M_{ej}$, then the period decrease due to the spin-up of the rotating NCE mass will exceed the period increase from the ejected mass. This percent of rotating mass to ejected mass where the orbital period decreases varies from 0.46\% at $q$ = 0.1 to 8.7\% at $q$ = 1. Therefore, the sudden decreases in the orbital periods across a classical nova outburst recently observed by \citep{sch20} can have a rational solution by including the spin-up of the NCE! It was recognized by \citet{kol01} that a large FAML could lead to an orbital period decrease.

The NCE continues to shrink to within the equipotential surface that intersects at the 2nd Lagrangian point and rotates with the binary system as it removes angular momentum from the binary. Fig.1d shows the secondary (red) and its convective region (dark purple) after it has re-accreted much more material from the NCE (light blue). Now, most of the NCE material that escapes the CV through the 2nd Lagrangian point flows by the secondary, which may capture some of it. This material will form a slow-velocity disk perpendicular to the binary's rotational axis behind the rapidly expanding nova shell. Recently, the observations of the nova V906 car \citep{mcl20} were best explained by a low-velocity circumbinary disk which most prominently appears 11 to 20 days following a classical nova outburst. The NCE ends when it shrinks to just within the WD's Roche lobe (Fig.1e). The remaining material in the WD's Roche lobe has its hydrogen burnt to helium in the remnant burning phase of a nova outburst.  This remnant burning is necessary to explain the excess helium abundance of the nova ejecta \citep{spa88}; the turn-off of X-rays observed from V1974 Cyg (Krautter et al. 1996), GQ Mus (Balman and Krautter 2001), and V745 Sco (Page et al. 2015); and the UV-line decline times for various novae (Vanlandingham et al. 2001). The average amount of remnant material, which seems to be common for all novae, on the WD is 25\% of the ejected mass \citep{pri89}. It is, therefore, reasonable that the secondary which collects material from its geometrical cross-section for the high-velocity ejected material, by its gravitational cross-section for the slower velocity material, from the NCE as it rotates within it, and from the rotating material that flows past the secondary when it escapes out the 2nd Lagrangian point can re-accrete similar amounts of nova-processed material. This amount is enough to increase the non-solar abundances of the secondary's surface layers \citep{ste99}.  In most of the above processes, the mass accreted on the secondary increases when its mass is larger.  We believe that the nova-processed material collected by the secondary during the NCE is the major source for the non-solar abundance CVs and has not been given adequate consideration. 

Eventually, the secondary overflows its Roche lobe (Fig.1f) again, only this time this material is mixed with nova-processed material. The inclusion of the NCE reveals how the non-solar abundances build up in the secondary's convective region. Obviously, the build-up of non-solar material on the secondary requires many nova outbursts. The amount of non-solar material in the secondary's convective region depends on its mass, the mass of the secondary when the secondary became a CV, the number of novae undergone, and the efficiency of the re-accretion. Provided the non-solar abundances from the NCE are more than the solar abundances from the increase in the secondary's convective region between novae outbursts, the non-solar abundances in the convectively mixed material will increase. As the secondary decreases in mass from its evolution, the convective region expands more rapidly between nova outbursts [see fig.14-2 in \citet{mac15}]. This means that, eventually, the abundances in the outer convective region become more solar-like. After the secondary becomes fully convective (at the top of the period gap) the non-solar abundances will again increase since there now is no dilution from the radiative core. Detailed model calculations \citep{ste99} give similar results. 

Lower mass WDs have weaker nova outbursts \citep{sta74} and probably more massive NCEs relative to their ejected mass. When the CV rotates within the NCE, FAML naturally occurs \citep{mac86}. Thus, this FAML will be more efficient in CVs with lower mass WDs than in those with more massive ones. Therefore, this FAML has the characteristics of the CAML required to bring calcuations into agreement with observations of the space density, orbital period distributions and the average WD mass of CVs \citep{sch16,bel18,zor19}. \citet{sch16} also speculated that their CAML might be caused from FAML following novae eruptions. In the commonly accepted scenario for CVs below the period gap, gravitational radiation is the only source for the loss of orbital angular momentum. Because this FAML results from a nova outburst, it is active throughout the evolution of the CV including above and below the gap. An additional source of angular momentum loss is needed to reconstruct the evolutionary path of CVs \citep{kni11} and re-produce the effective temperatures of their WDs \citep{pal17}. The period minimum problem can also be solved with an additional source of AML. Because our FAML depends on the amount of mass in the NCE and the mass of the secondary and because the mass in the NCE depends on the mass of the WD and its accretion rate, there will naturally be a large variation in the strength of this FAML among different CVs. This variation in the FAML strength will cause different decreases in the binary's separation leading to a spread in the secondary's Roche lobe mass overflow rate. This variation in the overflow rate can be the cause of the observed spread of the WD accretion rates at the same period as determined by the WD's effective temperatures. It can also be the cause of the nova-likes and DNs overlap over a long period range. \citet{kol01} noted that an additional angular momentum transfer from the binary orbit to the nova envelope from dynamical friction could lead to a decrease in the binary separation. Five of six classical novae are now known to have a sudden period decrease \citep{sch20} across their outburst contrary to expectations \citep{sha86}.

{\bf \section{Calculations}}

The following amplification to the standard nova outburst cycle in a CV demonstrates the role of the NCE in solving current CV problems. The rapid increase of energy during a TNR causes a strong shock wave to develop. As it propagates through the decreasing density layers, it leaves behind a large velocity gradient in the expanding envelope. Part of this material does not escape and forms an NCE. This NCE is strongly convective and extends beyond the secondary  \citep{spa76}. \citet{pri86} calculated that for a $1.25M_{\odot}$ CO WD this convective region lasted for 4.7 hr. and demonstrated that the mixing timescale was short enough for this region to be thoroughly mixed. The secondary sweeps up this nova-processed material which is mixed with its outer convective envelope. If the non-solar abundance is larger than the solar abundance from the expansion of the convective envelope between outbursts, the non-solar abundance will increase in this envelope. If the CV has undergone enough novae, these non-solar abundances will become observable. On the other hand, if the increase in solar abundance from the convective envelope expansion dominates, the non-solar abundance will decrease. This behavior was found by \citet{ste99} with calculations for a $1.0M_{\odot}$ WD and a $0.6M_{\odot}$ secondary. The non-solar abundances appearances (and disappearances) can be further explained  with Figure 14-2 from MacDonald (2015) for a single solar-abundance main-sequence star, which shows an increasing growth of its convective mass as the star's mass decreases until it becomes fully convective ($\sim0.3M_{\odot}$). This increase steepens at $\sim1.0M_{\odot}$.  All other things being equal, an initially more massive secondary will show surface re-accreted abundances sooner and longer in its evolutionary history. In the extreme case of a $1.4M_{\odot}$ secondary (and a $>1.4M_{\odot}$ WD), the convective region is only $1.5 \times 10^{-5} M_{\odot}$ deep which is approximately the amount of mass that this massive WD accretes before undergoing a TNR.  Thus, the secondary surface abundance will be enhanced by the re-accreted material of the previous outburst. As the mass of the secondary decreases from evolution, it will show re-accreted abundances on its surface until it is overwhelmed by the deepening of the convection region. The mass of the secondary when this happens is dependent on its initial mass when the binary becomes a CV, the number of novae undergone,  the composition of the NCE and the efficiency of the re-accretion. When the CV is below the period gap, the secondary is fully convective and there is no more radiative core to dilute it. Thus, its non-solar abundances will increase again.  Unfortunately, we do not know the evolutionary history of any CV. We can, however, make some general predictions. There should be a decrease of non-solar abundance CVs as the period approaches the period gap and then an increase below it. Table 1 seems to display this trend. However, because of the non-uniform studies yielding secondary's masses and abundances, and selection effects, we cannot confirm this by the secondary's observed mass. CVs with high secondary masses such as BV Cen, QZ Aur, BT Mon, V363 Aur and AC Cnc are good candidates for non-solar abundance characteristics.

Most studies \citep{sha86,mar97} on the accumulation of nova-processed material by the secondary have focused on the geometrical cross-section of the secondary.  \citet{mar98} mentioned the NCE together with the escaping nova shell but only included the latter in their study. \citet{ste99} found that for population I CVs, the accumulation must be an order of magnitude more than the geometrical cross-section. The nova-processed material as a source for the non-solar abundance should not be dismissed solely from this geometrical cross-section argument. Three-dimensional smoothed particle hydrodynamic (SPH) studies by \citet{fig18} finds that 3.0 to 3.6\% of the ejected mass is gravitationally captured by the secondary for a uniform high velocity (3000 km/s) compared to the  geometrical cross-section of 4.5\% for their model. However, for lower uniform velocities (800 and 1200 km/s) the amount of material captured increases by a factor of up to 3. Computational time requirements limit the models to $\sim 0.1 P$, and, therefore, does not cover the secondary interaction with the NCE.

We will now describe the changes in the composition of the outer convective region of the secondary due to material being re-accreted from the NCE and material from the extension into the radiative core. We follow the method of \citet{ste99} except we use a mass step (= the secondary's mass overflow rate times a time step) in the secondary's evolution instead of a time step, i.e., 

\begin{equation}
    M_{2}^{n+1} = M_{2}^{n} + f_{re} \Delta m_{of} - \Delta m_{of}
\end{equation}

where $M_{2}$ is the mass of the secondary, its superscript is its mass point in evolution, $\Delta m_{of}$ is the mass step, $f_{re}$ is the fraction of material re-accreted from the NCE. We simplify the detailed evolution \citep{ste99} of the secondary by estimating the mass of its outer convective region, $M_{2,C}^{n}$, from the convective mass-stellar relationship of \citet{mac15,mac20} at every mass step ($n$). The mass increase of the convective region over a mass step is then 

\begin{equation}
    \Delta m_{CI} = M_{2,C}^{n+1} - M_{2,C}^{n} + \Delta m_{of} - f_{re} \Delta m_{of}
\end{equation}

where $M_{2,C)}^{n}$ is the mass of the convective region corresponding to $M_{2}^{n}$.

Next,

\begin{equation}
M_{2,C,X_{i}}^{n+1} = M_{2,C,X_{i}}^{n} + f_{re} X_{NCE,i}\Delta m_{of} - X_{i}^{n+1/2}\Delta m_{of} + \Delta m_{CI} X_{sol,i},
\end{equation}

where $M_{2,C,X_{i}}^{n}$ is the mass of element $X_{i}$ in the convective region at mass step $n$, $X_{NCE,i}$ is its mass fraction in the NCE, $X_{sol,i}$ is its solar mass fraction and $X_{i}^{n+1/2}$ is its mass fraction loss averaged over the mass step. By replacing this last term with $X_{i}^{n}$, the explicit solution of the equation is found. The abundance of $X_{i}^{n+1}$ can be found by dividing the mass of the element in the convective region $M_{2,C,X_{i}}^{n+1}$ by the mass of the convective region $M_{2,C}^{n+1}$. $M_{1}$ (which is assumed to be constant), $M_{2}^{n=1}$ (the initial value of $M_{2}$ when the CV is born), $X_{sol,i}$, $X_{NCE,i}$, $f_{re}$ and $\Delta m_{of}$ are input parameters. $X_{i}^{n=1}$ is $X_{sol,i}$. Our model is evolved until $M_{2}^{n+1}$ decreases to its current mass. These simple calculations are repeated until $X_{i}$ matches the current observed value. Thus, the combination of the input values of $M_{2}^{n=1}$, $X_{NCE,i}$ and $f_{re}$ give a solution. We must investigate if these input values are viable. $M_{2}^{n=1}$ must be between its current value and $\le M_{1}$ for the normal CV evolution discussed earlier. At $M_{2}^{n=1}$ = $1.0M_{\odot}$ and above, there can be very little accumulation of material from the NCE in the secondary's convective region because it is approximately the amount of mass overflow needed for the next nova outburst. If $M_{2}^{n=1}$ is close to its current mass, $X_{i}$ can not change enough to be observed. $X_{NCE,i}$ must be a reasonable fraction of that element in the NCE, and $f_{re}$ is limited by a reasonable fraction of the mass in the NCE. Using the same input parameters except for $\Delta m_{of}$, the equations are tested to see if the results converge as $\Delta m_{of}$ decreases. It was found that convergence usually is reached if $\Delta m_{of} \le 0.01M_{\odot}$, and we use a value of $0.001 M_{\odot}$ in our calculations to insure convergence. In general, these equations yield the results described earlier in this section and by \citet{ste99} of non-solar abundance first increasing, then decreasing as the convective mass expands more rapidly into the radiative core until it becomes fully convective so that the non-solar abundances increase again. We also agree with \citet{ste99} that the non-solar abundances are not apparent unless the secondary's re-accretion is much more than its geometric cross-section indicates for population I CVs. 

With this simple but reasonable modeling of the secondary's evolution, we can rapidly explore reasonable combinations of the initial secondary's mass, the non-solar abundances in the NCE and the re-accreted mass from the NCE onto the secondary for {\it individual CVs}. N and C and their ratio N/C present additional constraints but a unique opportunity for this analysis. The calculated N/C ratio can be compared directly to the observed NV/CIV ratio. This ratio can vary between its original solar value (0.25) and its nuclear equilibrium value, which depends on the temperature in the burning region. For a nova outburst on a CO WD, C and N both come from the C originally on the WD, since O does not participate much in the nuclear burning at a maximum of $\sim$ 300 million degrees. Together C, N and O add up to most of the non-solar abundances in the CNO novae. Upon examining the observed compositions of novae ejecta gathered by \citet{sta98}, a large majority of the CNO novae have O within a factor of 2 of the sum of N and C. Since the N comes from proton capture on C, it is normally assumed that the CO WD's contribution to the nova ejecta is half C and half O when mixed with the accreted solar material in nova outburst studies. The total amount of CNO material observed in CNO nova ejecta \citep{sta98} ranges from 3.2\% to 57\%, and we must assume that the NCE material also has a wide range. We used values of 60\%, 40\%, 20\% and 10\% in our calculations which when divided equally between C  and O leads to initial C abundances of 30\%, 20\%, 10\% and 5\% in the NCE. \citet{pri86} demonstrated that the convection in the NCE was very rapid and its composition can be considered thoroughly mixed. Because of the different behaviors of the equilibrium N/C ratio in the fast and slow CN cycles and because we do not know the temperature in the burning region just before the convection ceases, we evolve our secondary models assuming that it is sweeping up material processed under burning temperatures associated with both the fast ($T_{6} \ge 50$) and the slow ($T_{6} \le 50$) CN cycle. The C's portion in the NCE is assumed to be split between C and N according to the assumed CN cycle. 

For V1309 Ori, \citet{sch01} measured emission line strengths of NV 1240 and CIV 1549 and found a ratio of 7.2, which we will use as the current N/C ratio in the secondary's convective region. We will first assume the cold CN cycle dominates just before convection ceases so that the equilibrium N/C ratio in the NCE is $\sim$ 100 \citep{cau62}. Since the WD's mass is 0.70$M_{\odot}$, we assume that it is a CO WD. We limit the secondary's initial mass by not exceeding the WD's mass but larger than its current mass of 0.5$M_{\odot}$. This mass range limitation and the rapid increase of the secondary's convective region at these masses results in the need for a large accumulation of nuclear-processed material for this CV. For an initial secondary mass of 0.70$M_{\odot}$ and an initial C abundance of 30\% in the NCE, a fractional re-accretion of 0.129 is required to reach a N/C ratio of 7.2 in the secondary's convective region. If the initial secondary's mass is 0.65$M_{\odot}$ or 0.6$M_{\odot}$, the corresponding re-accreted fraction is increased to 0.143 and 0.175, respectively. If the initial secondary's mass is 0.7$M_{\odot}$ and the initial C abundance is reduced to 20\%, the re-accretion fraction is increased to 0.182. In the hot CN cycle, the equilibrium C/N ratio is also dependent on the temperature, density ($\rho$), and hydrogen abundance ($X_{H}$) in the burning region. These additional parameters make the analysis more complex, but our simple secondary evolutionary models can still provide useful information. For a given $\rho X_{H}$, the equilibrium N/C first decreases and then increases with increasing temperature \citep{cau62}. By assuming the optimum conditions for re-accretion on the secondary, ie. the C composition of the NCE = 30\% and re-accretion rate = 0.3, we can determine the temperature range of the burning region at convection cutoff for each $\rho X_{H}$ where the observed value can not be reached. Using the optimum conditions to re-accrete and $\rho X_{H}$ = $10^{4}$, the temperature of the burning region at convection cutoff can not be between $T_{6}$ = 60 and 100. For $\rho X_{H}$ = $10^{3}$ and $10^{2}$, this temperature can not be between $T_{6}$ = 70 and 125 and $T_{6}$ = 90 and 175, respectively. Thus, the temperature of the burning region at convection turnoff must be lower than or greater than this range to match observations.
 
\citet{gan03} obtain a NV/CIV ratio of $>$ 14 from their HST spectrograph observations of EY Cyg. For the conditions of the cold CN cycle, an initial secondary's mass of 1.10$M_{\odot}$ equal to the (assumed to be a CO) WD's mass, an initial C abundance of 30\% in the NCE and a fractional re-accretion of 0.25, its N/C reaches a peak of 18.1 at $M_{2}$ = 0.82$M_{\odot}$ and decreases to 14.5 at its current mass of 0.49$M_{\odot}$. The initial secondary's mass can be decreased to 0.80$M_{\odot}$ with only a small reduction of N/C to 14.40. Therefore, we see that the results depend only weakly on the assumed initial secondary's mass if it is near 1.0$M_{\odot}$. When the fast CN cycle dominates, using the optimum conditions to re-accrete and $\rho X_{H}$ = $10^{4}$, $10^{3}$ and $10^{2}$, the temperature at convection turnoff can not be between $T_{6}$ = 50 and 120, $T_{6}$ = 50 and 160, and $T_{6}$ = 50 and 200, respectively. 

AE Aqr represents a different situation than EY Cyg because the initial secondary's mass is limited to $\le 0.63 M_{\odot}$ by the WD's mass where the secondary's convective mass is fairly large (0.099$M_{\odot}$) and increases to 0.316$M_{\odot}$ nearly the entire current secondary's mass (0.37$M_{\odot}$). To reach a ratio of $\sim$ 10 \citep{mau97}, an initial C abundance of 30\%, an initial secondary's mass of 0.63$M_{\odot}$ and a re-accretion rate of 0.23 is needed if the slow CN cycle dominates. When the fast CN cycle dominates, using the optimum conditions to re-accrete and $\rho X_{H}$ = $10^{4}$, $10^{3}$ and $10^{2}$, the temperature at convection turnoff can not be between $T_{6}$ = 50 and 110, $T_{6}$ = 50 and 140, and $T_{6}$ = 50 and 190, respectively.

\citet{mor02} found a mass ratio of 0.55$\pm$0.21 and lower limits of $0.48\pm0.32M_{\odot}$ for the secondary's mass and $0.87\pm0.24M_{\odot}$ for the WD's mass in GK Per. We have used the modeling of nova eruptions \citep{sha18} to obtain a WD mass of 1.22$M_{\odot}$ and the corresponding secondary's mass of 0.67$M_{\odot}$. These masses, which we will use for our calculations, fit within the constraints given by \citet{mor02}. The highest observed value of NV/CIV for GK Per is 1.6, which is easily obtained by a number of different input values. For an initial secondary's mass of 1.2$M_{\odot}$ and an initial C abundance of 30\%, the re-accretion fraction only has to be 0.01 for the N/C in the secondary's convective to reach 1.6. If the initial secondary's mass is reduced to just 0.7$M_{\odot}$, the re-accretion fraction is only raised to 0.05 for the N/C to reach 1.6. If the C abundance is reduced to 10\% and the initial secondary's mass is 1.2$M_{\odot}$, a re-accretion fraction of 0.03 again reaches 1.6. Even if the lowest equilibrium N/C of 4 for reasonable $\rho X_{H}$ values in the fast CN cycle is used there are no additional limitations.

At the lower period end of table 1, we find EI Psc (RX J2329) well below the period gap with a WD mass of 0.65$M_{\odot}$ and a secondary mass of 0.13 $M_{\odot}$. \citet{gan03} find a NV/CIV ratio of $>$ 9. As long as the initial mass of the secondary is not much smaller than where the secondary becomes fully convective, the input values to reach N/C $>$ 9 are fairly modest because there is no radiative core to dilute the re-accreting material. If the initial secondary's mass is 0.35$M_{\odot}$ and the C abundance is 20\%, the re-accretion fraction is 0.16 to reach a N/C $>$ 9 and when the C abundance is reduced to 10\%, the re-accretion must be increased to 0.25. Raising the initial secondary's mass to 0.65$M_{\odot}$ reduces the re-accretion fraction to 0.18. Again when the fast CN cycle dominates, using the optimum conditions to re-accrete and $\rho X_{H}$ = $10^{4}$, $10^{3}$ and $10^{2}$, the temperature at convection turnoff can not be between $T_{6}$ = 60 and 100 degrees, $T_{6}$ = 80 and 125 degrees, and $T_{6}$ = 90 and 170 degrees, respectively. 

V396 Hya (CE 315) and GP Com also produce large NV/CIV line ratios and are below the period gap, but are AM CVn CVs (double degenerate dwarfs). These CVs contain little or no hydrogen and are not applicable to our scenario [see Nelemans et al. (2010) for a discussion of possible evolutionary paths]. TX Col, MN Hya, BY Cam, BZ UMa and 1108+5728 (= CSS 120422:111127+571239) have NV/CIV values but no published masses. The primary mass (1.95$M_{\odot}$) of CH UMa is too large to be a WD and the secondary mass (3.32$M_{\odot}$) of V Sge is far outside the normal range of CVs. Studies of the NV/CIV ratio also include CVs with normal solar ratios. In all of the above cases, the calculated N/C ratio can easily match the observations if the CV is population II. Table 1 shows CVs with a solar NV/CIV ratio (indicated by a \enquote{N}) become more dominate as the binary period decreases until the cluster of abnormal NV/CIV below the gap. 

Most designations of nonsolar $^{12}\text{C}/^{13}\text{C}$ are due to the appearance of $^{13}$CO absorption where none is expected. Numerical values of $^{13}\text{C}/^{12}\text{C}$ for AE Aqr, SS Cyg and RU Peg are found in \citet{har17}. AE Aqr and SS Cyg have $^{12}\text{C}/^{13}\text{C} \sim$ 4.  With the current observed secondary masses of 0.37$M_{\odot}$ and 0.55$M_{\odot}$, respectively, and a burning temperature at convection turnoff $T_{6}<$ 50, ie. the slow CN cycle, it is not possible to achive such a low ratio because the convective region is expanding into the radiative core too rapidly. For a temperature of $T_{6}$ = 120 and the fast CN cycle, \citet{cau62} found the equilibrium abundance ratios of N/C and $\text{C}^{12}/\text{C}^{13}$ for different values of $\rho X_{H}$. With our evolutionary model of the secondary, it is possible to find a number of conditions ($\rho X_{H}$, $M_{2,init}$, re-accretion fraction and initial C abundance), where the composition of the secondary's convection matches the observations of both AE Aqr and SS Cyg. Other equilibrium ratios are not given by \citet{cau62} except for T$_{6}$ = 300 which is too high for our burning temperature. However, the equilibrium $^{12}\text{C}/^{13}\text{C}$ ratios that they found for minimum equilibrium N/C values imply that values of $^{12}\text{C}/^{13}\text{C}$ at other temperatures can match observations. All of this boils down to the temperature $T_{6}$ must be $>$50 at convection turnoff in AE Aqr and SS Cyg to match observations. 
If AE Aqu and SS Cyg are population II CVs their $^{12}\text{C}/^{13}\text{C}$ ratios can be this low with the slow CN cycle. \citet{ste99} found that for population II CVs, the abundance of the secondary's convective region was dominated by nova ejected material. 
\citet{har17}'s preferred value of $^{12}\text{C}/^{13}\text{C}$ = 32 for RU Peg presents an interesting case. Its secondary mass is 0.96$M_{\odot}$ so that its present convective region is small, and its primary mass is 1.06$M_{\odot}$, which limits the range of the initial mass of the secondary. For the slow CN cycle and an initial secondary mass of 1.06$M_{\odot}$, combinations of the C envelope abundance and accreted fraction of the envelope of (10\%, 0.33), (20\%, 0.2) and (30\%, 0.14) give $^{12}\text{C}/^{13}\text{C}$ ratios of $\sim$32. Decreasing the initial secondary mass to 1.01$M_{\odot}$ only raises this ratio to $\sim$34. When the fast CN cycle dominates at convection turnoff, The observed $^{12}\text{C}/^{13}\text{C}$ can easily be reached because of the much lower equilibrium $^{12}\text{C}/^{13}\text{C}$ values.

\newpage{}
\vspace{5mm}
\section{Conclusions and Discussion}

The shock wave from a TNR in the WD of a CV will naturally produce a NCE. As the CV rotates within the NCE, it will sweep part of it up, eject part of it and force part to co-rotate with it. The part swept up by the secondary is a viable source of non-solar material for it. The CV's interaction with the other two parts innately removes angular momentum from the CV binary via FAML. This FAML should be stronger in CVs with lower periods which is characteristic of the CAML needed for the average CV WD mass, the space density and the orbital period problems. Since this FAML continues when the CVs are below the period gap, it may also be the CAML needed for the minimum period problem and the WD effective temperature issue. The effectiveness of this FAML will no doubt be dependent on the strength of the nova outburst, the individual component masses and their separation amongst other factors. These variables can lead to a larger scatter of the mass transfer rates which is observed but not predicted \citep{kol01} and of the WD effective temperatures near period minimum \citep{pal17}. This FAML will cause the orbital period to decrease. This period decrease may be large enough to overcome the increase of the orbital period from the nova ejected material and explain the sudden {\it decrease} in the orbital period of a CV across a nova outburst \citep{sch20}. This FAML is {\it only} effective during the outburst of the CV's nova cycle. This can explain the {\it sudden} decrease in the orbital period. The NCE's influence on the abundances of the secondary and the evolution of the CV becomes effective only when the pre-CV becomes a CV and continues during its whole lifetime. This behavior is compatible with the different average WD mass of the pre-CVs, CVs and single stars, the non-solar abundances of some CVs but not the pre-CVs, and the additional angular momentum loss below the period gap.

Table 1 reveals a decrease in the number of CVs with non-solar abundances as the period decreases down to the period gap and then an increase below the period gap. Because of the rapid increase in the secondary's convective region with the decrease in the secondary's mass until the radiative core is gone, this is the expected behavior of our proposed mechanism for the non-solar abundances in the CV's secondaries. The decrease in the secondary's capacity ($\gamma$) with the mass ratio and the amount of re-accreted material decrease as the secondary's mass decreases for our mechanism also agrees with this tabular trend. The EMS scenario does not perdict this behavior and the red giant CE phase does not predict the increase after the period gap. AE Aqr is unique among the CVs with non-solar abundances in that both $^{12}\text{C}/^{13}\text{C}$ and NV/CIV have numerical values. The calculated $^{12}\text{C}/^{13}\text{C}$ ratio can not match observations if the slow CN cycle dominates at convection turnoff, ie, $T_{6} < 50$. For the fast CN cycle the temperature at convection turnoff must then be $T_{6} > 110$ to match its observations of NV/CIV. This implies that the convection in the NCE turned off early because the temperature in the burning region decreases rapidly after its peak and provides a constraint for future nova modeling of AE Aqr. As more abundance data is collected, we will be able to provide constraints on the modeling of individual novae and their NCEs.

\citet{har05b} raised the question of why there are 
no reductions or absences of the infrared CO absorption feature in the pre- or magnetic CVs. The lack of non-solar abundance CVs in the pre-CVs is a natural result of the NCE being the source. Since this question was raised, some longer period magnetic CVs do show enhanced NV/CIV ratios \citep{gan03}. It is not clear if the shorter period magnetic CV have less non-solar abundance characteristics or not. If they do, then a possible explanation is that since the WD's magnetic field controls the flow of material from the secondary, it can redirect some of the NCE material that the secondary would have collected onto the WD instead. For the case of magnetic CVs with non-solar abundances far above the gap, their long period and large binary separation will reduce the effectiveness of the WD's magnetic field shielding the secondary.

Both the NCE's influence on the abundances of the secondary and the evolution of the CV becomes effective, when the pre-CV becomes a CV and continues during its whole lifetime. This behavior is compatible with the different characteristics of the pre-CVs and CVs, the non-solar abundances of some CV but not the pre-CVs and the scaling up of the angular momentum loss due to gravitational radiation below the period gap. The fact that both the non-solar abundances and the large average WD mass occur in CVs but not in pre-CVs could be an indication that their root cause is the same.  That no pre-CVs \citep{how10} show abundance anomalies is a natural result of our NCE theory. This is a major distinction over the CE phase and EMS scenario.  Breakout from the CNO cycle to produce nuclei like Ne, Na, Mg and Ar requires temperatures above 300 million degrees \citep{gil03}. These non-solar abundances are a problem for the EMS scenario and the CE phase, which is easily solved when the cause is a nova outburst \citep{sio14}. The high over-abundance of Argon in U Gem \citep{Go17} can {\it only} come from an ONeMg WD.  Thus, using the average mass accretion onto its WD for U Gem from Dubus et al. (2018) and the accreted mass needed for a TNR on a 1.25 $M_{\odot}$ ONeMg WD from Politano (1995), we can predict that U Gem will erupt as a ONeMg nova within 45,000 yrs. The high abundances of odd-numbered nuclear species above oxygen, which must come from high temperature H-capture on even numbered species, and the extreme deficits of carbon for EI Psc and QZ Ser \citep{har16,har17} also makes them candidates for having ONeMg WDs. In fact, once the analysis is accurate enough to distinguish the abundances of these metals on many CVs, the number of ONeMg WDs relative to the CNO WDs can be determined {\it directly}. The present determination involves an assumption of the outburst strengths of the two different types of novae \citep{gil03}.  This ratio is of fundamental importance to the evolution of massive stars and the study of novae.

\section{Acknowledgements}

This work is supported by NASA ADP grant NNX13AF12G and NSF grant AST-1413537 to Villanova University. We wish to thank Jim MacDonald for providing his numerical table of convective mass versus secondary mass, Ron Webbink for discussions on accretion onto convecting secondaries, Austin Leithner for writing the JAVA program for the secondary's evolution, and the referee for pointing out shortfalls and errors.

\begin{table}
\caption{Cataclysmic Variables with known Periods and Mass Ratios or Non-Solar Abundance Indicators}
\begin{center}
\begin{tabular}{lllcccccccc}
\tableline 
\noalign{\smallskip}
System & 	CV   & $P_{orb}$    & q       & $M_2$ & NV/CIV & 13C/12C & Near-IR & IR cont & CO & UV \\  
Name & 	type & $<$d$>$ &$<M_2/M_1>$& $<M_{\odot}>$     &     &      & lines &      &        &    \\  
\tableline 
\noalign{\smallskip}
V1017 Sgr   &	Nb    &	5.786038     &	0.5   &	0.6   &     &       &       &       &       &       \\ 
V603 Cas    &	DN    &	2.56387	     &	0.183 &	0.18  &     &       &       &       &       &       \\ 
GK Per	    &   Na IP &	1.996803     &	0.55  &	0.172 &	V   &	    &	Y   &	N   &	Y   &       \\ 
J2044-0459  &	DN UG &	1.68	     &	0.36  &       &     &       &       &       &       &       \\ 
U Sco	    &	Nr    &	1.230547     &	0.55  & 0.88  &     &       &       &       &       &       \\ 
V508 Dra    & 	      &	0.838042     &	      &	      &	    &	    &	    &       &	Y   &       \\ 
CI Aql	    &	Nr    &	0.618363     &	2.35  &	2.32  &     &       &       &       &       &       \\ 
BV Cen	    &	DN UG &	0.611179     &	0.89  &	1.05  &	    &       &       &       &       &       \\ 
J1930+0530  &	CV?   &	0.61092	     &	1.78  &       &	    &	    &	    &       &       &       \\ 
V Sge	    &	SS    &	0.514197     &	3.76  & 3.32  &	V   &       &       &       &       &       \\ 
UY Pup      &	DN UG &	0.479269     &	0.91  &	      &	    &	    &	    &       &       &       \\ 
EY Cyg	    &   DN UG &	0.459324     &	0.44  &	0.49  &	Y   &	    &	Y   &	Y   &	Y   &       \\ 
DX And	    &	DN UG &	0.440502     &	0.96  &	      &	    &	    &	Y   &   Y   &       &       \\ 
AE Aqr	    &	NL DQ &	0.411656     &	0.6   &	0.37  &	Y   &	Y   &	Y   &	Y   &	Y   &       \\ 
NY Lup	    &	NL IP &	0.411	     &	0.65  & 0.6   &     &       &       &       &       &       \\ 
SY Cnc	    &	DN ZC &	0.382375     &	0.68  &	      &	    &	    &	N   &	N   &	    &       \\ 
AT Ara	    &	DN UG &	0.3755	     &	0.79  &	0.42  &     &       &       &       &       &       \\ 
RU Peg	    &	DN UG &	0.3746       &	0.878 & 0.96  & N   &	Y   &	Y   &	N   &	Y   &	Y   \\ 
QZ Aur	    &	Na    &	0.35749703   &	0.95  & 0.93  &     &       &       &       &       &       \\ 
CH UMa	    &	DN UG &	0.343184     &	0.49  & 0.96  & Y   &	Y   &	Y   &	    &	Y   &       \\ 
MU Cen	    &	DN UG &	0.342	     &	0.83  &	0.99  &	    &	Y   &	    &	    &	Y   &       \\ 
BT Mon	    &	Na NL &	0.33381490   &	0.84  & 0.87  &	    &	    &	    &	    &	    &       \\ 
V1309 Ori   &	NL AM &	0.332612     &	0.67  & 0.5   & Y   &	    &	Y   &	    &	Y   &       \\ 
V392 Hya    &	DN UG?&	0.324952     &	0.55  &       &     &       &       &       &       &       \\ 
V363 Aur    &	NL SW &	0.321242     &	1.17  &	1.06  &     &       &       &       &       &       \\ 
J0345+5335  &	CV DN?&	0.3139	     &	0.83  &       &     &       &       &       &       &       \\ 
RY Ser	    &	DN    &	0.3009	     &	0.8   &       &     &       &       &       &       &       \\ 
AC Cnc	    &	NL SW &	0.300478     &	1.02  &	0.77  &	    &	    &	    &	    &	    &       \\ 
V838 Her    &	Na    &	0.297635     &	      &	0.74  &	    &	    &	    &       &       &       \\ 
0218+3229   &	DN    &	0.29723	     &	0.58  &	0.34  &     &       &       &       &       &       \\ 
EM Cyg	    &	DN ZC &	0.290909     &	0.77  &	0.77  &	    &	Y   &	Y   &	    &	    &       \\ 
Z Cam	    &	DN ZC &	0.289841     &	0.71  & 0.70  &     &	    &	N   &	N   &       &       \\ 
V426 Oph    &	DN IP?&	0.2853	     &	0.78  & 0.70  &	    &	N   &	    &	    &	N   &       \\ 
SS Cyg 	    &	DN UG &	0.27513	     &	0.685 & 0.55  &	N   &	Y   &	YH  &	N   &	Y   &       \\ 
BF Eri	    &	DN NL &	0.27088	     &	0.41  & 0.52  &     &       &       &       &       &       \\ 
CW 1045+525 &	DN    &	0.271278     &	      &       & Y   &       &       &       &       &       \\ 
J0644+3344  & 	NL SW?&	0.269374     &	0.778 & 0.52  &     &       &       &       &       &       \\ 
TT Crt	    &	DN UG &	0.26842	     &	      &	      &	    &	    &	Y   &       &       &       \\ 
AH Her	    &	DN ZC &	0.258116     &	0.8   & 0.76  &	    &	Y   &	N   &	N   &	Y   &       \\ 
J1931+4559  &	NL DN?& 0.2544	     &	0.83  &       &     &       &       &       &       &       \\ 
XY Ari 	    &	NL IP &	0.252697     &	0.58  & 0.62  &     &       &       &       &       &       \\ 
J1544+2553  &	CV    &	0.25128168   &	0.52  &       &     &       &       &       &       &       \\ 
RW Sex	    &	NL UX &	0.2451450    &	0.74  & 0.674 &     &       &       &       &       &       \\ 
TX Col	    &	NL IP &	0.2383	     &	      &	      &	Y   &       &       &       &       &       \\ 
J1744-2603  &	DN?   &	0.237089     &	0.75  &       &     &       &       &       &       &       \\ 
VY Scl	    &	NL VY &	0.2323	     &	0.32  &       &     &       &       &       &       &       \\ 
V347 Pup    &	NL SW &	0.231936     &	0.83  &  0.52 &	N   &       &       &       &       &       \\ 
RW Tri 	    &	NL SW &	0.2318833    &	0.60  &  0.42 &	N   &       &       &       &       &       \\ 
TV Col	    &	NL IP &	0.2286       &        &	 0.56 &	N   &       &       &       &       &       \\ 

\tableline 
\end{tabular} 
\end{center}
\end{table} 

\begin{table}[]
\begin{center}
\begin{tabular}{lllcccccccc}
\tableline 
\noalign{\smallskip}
System & 	CV   & $P_{orb}$    & q       & $M_2$ & NV/CIV & 13C/12C & Near-IR & IR cont & CO & UV \\  
Name & 	type & $<$d$>$ &$<M_2/M_1>$& $<M_{\odot}>$     &     &      & lines &      &        &    \\  
\tableline 
\noalign{\smallskip}

CZ Ori	    &	DN UG &	0.2189	     &	      &	      &     &	    &	    &	N   &       &       \\ 
HL CMa 	    &	DN ZC &	0.216787     &        &       &     &       &       &       &       &       \\ 
HR Del	    &	Nb    &	0.21416215   &	0.82  &  0.55 &     &       &       &       &       &       \\ 
EX Dra	    & 	DN UG &	0.209937     &	0.75  &	 0.56 &	    &	    &	Y   &	    &	N   &       \\ 
RX And	    &	DN ZC &	0.209893     &	0.42  &	 0.48 &	    &	    &	N   &       &       &       \\ 
T Aur	    &	Nb    &	0.204378     &        &	 0.63 &     &       &       &       &       &       \\ 
FO Aqr	    &	NL IP &	0.202060     &	      &	      &	N   &       &       &       &       &       \\ 
UX UMa	    &	NL UX &	0.196671     &	0.43  &  0.39 &	N   &       &       &       &       &       \\ 
IX Vel	    &	NL UX &	0.193927     &	0.65  &  0.53 &     &       &       &       &       &       \\ 
DQ Her	    &	Na IP &	0.1936208977 &	0.67  &  0.40 &	N   &       &       &       &       &       \\ 
J0107+4845  &	NL SW &	0.193598     &	0.429 &       &     &       &       &       &       &       \\ 
J1006+2337  &	DN UG &	0.185913     &	0.51  &  0.41 &     &       &       &       &       &       \\ 
MQ Dra	    &	NL AM &	0.18297	     &	      &	      &	    &	    &	N   &	    &	N   &       \\ 
SS Aur	    &	DN UG &	0.1828	     &	0.36  &	 0.39 &	    &	    &	N   &	    &	N   &   N   \\ 
TW Vir	    &	DN UG &	0.182682	     &	0.44  &	 0.40 &	    &	    &	    &	    &	Y   &       \\ 
BD Pav	    &	DN UG &	0.179301     &	0.63  &  0.73 &     &       &       &       &       &       \\ 
U Gem	    &	DN UG &	0.176906     &	0.35  &  0.42 &	    &	    &	Y   &	    &	Y   &	YH  \\ 
CW Mon	    &	DN UG & 0.1766	     &	      &	      &	    &	    &	N   &       &       &       \\ 
WW Cet	    &	DN ZC &	0.1758	     &	0.49  &	 0.41 &	    &	    &	Y   &	    &	    &	Y   \\ 
GY Cnc	    &	DN UG &	0.175442     &	0.448 &	 0.38 &     &       &       &       &       &       \\ 
V1043 Cen   &	NL AM &	0.174592     &  0.45  &       &     &       &       &       &       &       \\ 
J2216+4646  &	XL IP &	0.171802     &  0.83  &  0.45 &     &       &       &       &       &       \\ 
YY/DO Dra   &	DN IP &	0.165374     &	0.45  &	 0.38 &	    &	    &	Y   &	    &	    &	Y   \\ 
J1927+4447  &	DN    &	0.165308     &	0.57  &  0.39 &     &       &       &       &       &       \\ 
V1776 Cyg   &	NL UX &	0.164739     &        &  0.37 &     &       &       &       &       &       \\ 
UU Aqr	    &	NL UX &	0.163805     &	0.30  &  0.20 &     &       &       &       &       &       \\ 
UU Aql	    &	DN UG &	0.1635324    &	      &	      &	    &	    &	    &	    &   Y   &       \\ 
CN Ori	    &	DN UG &	0.163199     &	0.66  &	 0.49 &     &       &       &       &       &       \\ 
KR Aur 	    &	NL VY &	0.1628	     &	0.39  &	 0.37 &     &       &       &       &       &       \\ 
KT Per	    &	DN UG &	0.162656     & 	      &       &	    &	    &	N   &	    &	    &	N   \\ 
CM Del	    &	NL UX &	0.162	     &	0.75  &  0.36 &     &       &       &       &       &       \\ 
LX Ser 	    &	NL SW &	0.158432     &	0.88  &  0.36 &     &       &       &       &       &       \\ 
IP Peg	    &	DN UG &	0.158206     &	0.48  &	 0.55 &	    &       &	N   &	    &	N   &	N   \\ 
OY Ara	    &	Na SW &	0.155390     &  0.41  &  0.34 &     &       &       &       &       &       \\ 
GS Pav      &	NL VY &	0.155270     &	      &       & Y   &       &       &       &       &       \\ 
QQ Vul	    &	NL AM &	0.15452	     &	0.63  &  0.42 &	N   &       &       &       &       &       \\ 
V380 Oph    &	NL SW &	0.1534766    &	0.62  &  0.36 &     &       &       &       &       &       \\ 
QS Vir	    &	DS    &	0.150758     &	0.52  &	      &	    &	    &	    &	    &	N   &       \\ 
V425 Cas    &	NL VY &	0.1496	     &	0.36  &  0.31 &     &       &       &       &       &       \\ 
0220+0603   &	NL SW &	0.149208     &	0.54  &  0.47 &     &       &       &       &       &       \\ 
J0537-7034  &	      &	0.147	     &	      &	 0.35 &     &       &       &       &       &       \\ 
RR Pic	    &	Nb SW &	0.1450237620 &	0.4   &	 0.4  &     &       &	    &	    &	    &	Y   \\
J1007-2017  &	NL AM &	0.144864     &	      &	 0.35 &     &       &       &       &       &       \\ 
VZ Scl	    &	NL VY &	0.144622     &	      &	 0.32 &     &       &       &       &       &       \\ 
MN Hya	    &	NL IP &	0.141244     &	      &	      &	Y   &       &       &       &       &       \\  
V1223 Sgr   &	NL IP &	0.140244     &	      &	 0.40 &     &       &       &       &       &       \\ 
V1432 Aql   &	NL AM &	0.140236     &	      &	      & N   &       &       &       &       &       \\ 
BY Cam	    &	NL AM &	0.139753     &	      &	      &	Y   &	    &	    &	    &	    &	Y   \\ 
V1315 Aql   &	NL SW &	0.139690     &	0.41  &  0.30 &     &       &       &       &       &       \\ 
V728 Sco    &	Na NS &	0.13833866   &	0.36  &  0.29 &     &       &       &       &       &       \\ 
V603 Aql    &	Na NS &	0.138201     &	0.24  &	 0.29 &     &       &       &       &       &       \\ 
J0756+0858  &	NL SU &	0.136975     &	0.47  &  0.28 &     &       &       &       &       &       \\ 
DW UMa	    &	NL SW &	0.136607     &	0.28  &  0.30 &     &       &       &       &       &       \\ 

\tableline 
\end{tabular} 
\end{center}
\end{table} 

\begin{table}[]
\begin{center}
\begin{tabular}{lllcccccccc}
\tableline 
\noalign{\smallskip}
System & 	CV   & $P_{orb}$    & q       & $M_2$ & NV/CIV & 13C/12C & Near-IR & IR cont & CO & UV \\  
Name & 	type & $<$d$>$ &$<M_2/M_1>$& $<M_{\odot}>$     &     &      & lines &      &        &    \\  
\tableline 
\noalign{\smallskip}

BG CMi	    &	NL IP &	0.134748     &        &  0.38 &     &       &       &       &       &       \\ 
J0837+3830  &	NL AM &	0.1325	     &	      &	 0.30 &	    &	    &       &       &   N   &       \\ 
MV Lyr	    &	NL VY &	0.132335     &	0.43  &  0.30 &     &       &       &       &       &       \\ 
AM Her	    &	NL AM &	0.128927     &        &       &	N   &	    &	N   &	    &	N   &       \\ 
J0545+0221  &	DN    &	0.120972     &	0.236 &       &     &       &       &       &       &       \\ 
HY Eri	    &	NL AM &	0.118969     &	0.83  &  0.36 &     &       &       &       &       &       \\ 
TU Men	    &	DN SU &	0.1172       &	      &	      &     &       &       &       &   Y   &       \\ 
V348 Pup    &	NL SW &	0.101839     &	0.253 &  0.19 &     &       &       &       &       &       \\ 
J0011-0647  &	DN UG &	0.100281     &	0.32  &  0.33 &     &       &       &       &       &       \\ 
V1239 Her   &	DN SU &	0.100082     &	0.248 & 0.223 &     &       &       &       &       &       \\ 
J0350+3232  &	NL AM &	0.0988223    &	0.211 &       &     &       &       &       &       &       \\ 
J0750+1411  &	DN?   &	0.093164     &	0.59  &       &     &       &       &       &       &       \\ 
V1258 Cen   &	DN SU &	0.088941     &	0.246 &	 0.18 &     &       &       &       &       &       \\ 
IR Com	    &	DN?   &	0.087039     &	      &	 0.18 &     &       &       &       &       &       \\ 
HU Aqr	    &	NL AM &	0.0868204    &	0.22  &       &     &       &       &       &       &       \\ 
YZ Cnc	    &	DN SU &	0.0868	     &	0.222 &	 0.17 &     &       &       &       &       &       \\ 
DV UMa	    &	DN SU &	0.085853     &	0.178 & 0.196 &     &       &       &       &       &       \\ 
EF Peg	    &	DN SU &	0.0837	     &	0.26  &       &     &       &       &       &       &       \\ 
QZ Ser	    &	DN SU &	0.08316078   &	      &	      &	    &	Y   &	YH  &	Y   &	Y   &       \\ 
AR UMa	    &	NL AM &	0.0805006    &	      &	      &	    &	    &	N   &	    &	N   &       \\ 
WW Hor	    &	NL AM &	0.080199     &	0.18  &	 0.19 &     &       &       &       &       &       \\ 
ST LMi	    &	NL AM &	0.079089     &	0.22  &	 0.17 &	N   &	    &	N   &	    &	N   &       \\ 
V2301 Oph   &	NL AM &	0.07845	     &	0.15  &       &     &       &       &       &       &       \\ 
CU Vel	    &	DN SU &	0.078054     &	0.115 &  0.15 &     &       &       &       &       &       \\ 
MR Ser	    &	NL AM &	0.078798     &	      &	      &	    &	    &	N   &	    &	N   &       \\ 
SU UMa	    & 	DN SU &	0.07637540   &	      &	      &	N   & 	    &	    &	    &	    &       \\ 
RZ Leo	    &	DN SU &	0.076038     &	      &	      &	    &	    &	N   &	    &	N   &       \\ 
V893 Sco    &	DN SU &	0.075961     &	0.19  &  0.175 &    &       &       &       &       &       \\ 
WX Hyi	    &	DN SU &	0.074813     &	0.18  &  0.16 &	N   &       &       &       &       &       \\ 
Z Cha	    &	DN SU &	0.074499     &	0.189 &  0.125&	N   &	    &	    &	    &	Y   &       \\ 
VW Hyi	    &	DN SU &	0.074271     &	0.147 &  0.11 &	    &	    &	N   &	    &	N   &	YH  \\
IY UMa	    &	DN SU &	0.073909     &	0.146 &  0.10 &	    &       &       &       &       &       \\ 
HT Cas	    &	DN SU &	0.073647     &	0.15  &  0.09 &     &       &       &       &       &       \\ 
OU Vir	    &	DN SU &	0.072706     &	0.1641 & 0.116 &    &       &       &       &       &       \\ 
EP Dra	    &	NL AM &	0.072656     &	0.31  &  0.133 &    &       &       &       &       &       \\ 
V834 Cen    &	NL AM &	0.070498     &	      &       &	N   &       &       &       &       &       \\ 
V2779 Oph   &	DN SU?&	0.070037     &	0.168 &       &     &       &       &       &       &       \\ 
VV Pup	    &	NL AM &	0.069747     &   0.15 &  0.10 &	    &	    &	N   &	    &	N   &       \\ 
EX Hya 	    &	NL IP &	0.068234     &	0.128 &  0.108&	N   &	    &	Y   &	    &	N   &       \\ 
HS Cam	    &	NL AM &	0.068207     &	0.175 &  0.13 &     &       &       &       &       &       \\ 
BZ UMa	    &	DN IP?&	0.06799	     &	      &	      &	Y   &	    &	    &	    &   Y   &       \\ 
J1152+4049  &	NL SU &	0.0676	     &	0.155 &  0.087&     &       &       &       &       &       \\ 
BS Tri	    &	DN SU &	0.066881     &	0.21  &  0.16 &     &       &       &       &       &       \\ 
DI Phe	    &	DN    &	0.06555	     &	0.1097 & 0.101&     &       &       &       &       &       \\ 
J1524+2209  &	DN SU &	0.065319     &	0.17  &       &     &       &       &       &       &       \\ 
QZ Lib	    &	DN SU &	0.06436	     &	0.040  &      &     &       &       &       &       &       \\ 
OY Car	    &	DN SU &	0.063121     &	0.1065 & 0.086&	N   &       &       &       &       &       \\ 
VY Aqr	    &	DN WZ &	0.06309	     &	0.0660 &      &	    &	    &	Y   &	    &	Y   &	N   \\
GP CVn	    &	DN SU &	0.06295	     &	0.1115 & 0.089&     &       &       &       &       &       \\ 
EK TrA 	    & 	DN SU &	0.06288	     &	0.20   & 0.09 &     &       &       &       &       &       \\ 
BC UMa	    & 	DN WZ &	0.06261	     &	       &      &	    &	    &	    &	    &	    &	Y   \\ 
V436 Cen    & 	DN SU &	0.062501     &	0.24   & 0.17 &	    &	    &	N   &	    &	N   &       \\ 
V2051 Oph   & 	DN SU &	0.062428     &	0.19   & 0.15 &	    &	    &	N   &	    &	N   &       \\ 
DP Leo	    & 	NL AM &	0.062363     &	0.125  & 0.14 &     &       &       &       &       &       \\ 

\tableline 
\end{tabular} 
\end{center}
\end{table} 

\begin{table}[]
\begin{center}
\begin{tabular}{lllcccccccc}
\tableline 
\noalign{\smallskip}
System & 	CV   & $P_{orb}$    & q       & $M_2$ & NV/CIV & 13C/12C & Near-IR & IR cont & CO & UV \\  
Name & 	type & $<$d$>$ &$<M_2/M_1>$& $<M_{\odot}>$     &     &      & lines &      &        &    \\  
\tableline 
\noalign{\smallskip}

V4140 Sgr   & 	DN SU &	0.061430     &	0.125  & 0.09 &     &       &       &       &       &       \\ 
XZ Eri	    & 	DN SU &	0.061159     &	0.1098 & 0.091 &    &       &       &       &       &       \\ 
J0903+3300  & 	DN SU &	0.059074     &	0.113  & 0.099 &    &       &       &       &       &       \\ 
NZ Boo	    &	DN SU &	0.058910     &	0.1099 & 0.078 &    &       &       &       &       &       \\ 
QZ Vir	    &	DN IP?&	0.05882047   &	0.70   & 0.11 &     &       &       &       &       &       \\ 
EG Cnc	    &	DN WZ &	0.05877	     &	       &      &	    &	    &	    &	    &	    &	N   \\ 
CC Scl	    &	DN IP &	0.058567     &  0.072  &      &     &       &       &       &       &       \\ 
WX Cet	    &	DN WZ &	0.058261     &	0.085  & 0.047&	    &	    &	    &	    &	    &	N   \\ 
HV Vir	    &	DN WZ &	0.057069     &	       &      &	    &	    &	    &	    &	    &	N   \\ 
J1035+0551  &	DN? BD&	0.057007     &	0.057  & 0.048&     &       &       &       &       &       \\ 
V1838 Aql   &	DN SU &	0.05698      &	0.10   &      &     &       &       &       &       &       \\ 
J1501+5501  &	DN SU &	0.056841     &	0.101  & 0.077&     &       &       &       &       &       \\ 
SW UMa	    &	DN SU &	0.056815     &	0.14   & 0.010&     &	    &	    &	    &	    &	Y   \\ 
WZ Sge	    &	SU WZ &	0.056688     &	0.092  & 0.078&	    &	    &	    &	    &	    &	N   \\ 
AL Com	    &	DN WZ &	0.056669     &	       &      &	    &	    &	    &	    &	    &	N   \\ 
EF Eri	    &	NL AM &	0.056266     &	      &	      &	    &	    &	    &	    &	Y   &       \\ 
GG Leo	    &	NL AM &	0.055471     &	0.12  &	 0.09 &     &       &       &       &       &       \\ 
LL And	    &	DN WZ &	0.055055     &	      &	      &	    &	    &	    &	    &	    &   N   \\ 
V627 Peg    &	DN WZ &	0.054523     &	0.169 &  0.13 &     &       &       &       &       &       \\ 
BW Scl	    &	DN WZ &	0.054323     &	      &	      &	    &	    &	    &	    &	    &	Y   \\ 
J1433+1011  &	DN    &	0.054241     &	0.066 &	 0.057&     &       &       &       &       &       \\ 
GW Lib	    &	DN WZ &	0.05332	     &	0.060 &	 0.050&	    &	    &	    &	    &	    &	N   \\ 
KN Cet	    &	DN? BD&	0.052985     &	0.088 &	 0.064&     &       &       &       &       &       \\ 
J1944+4912  &	DN SU &	0.052816     &	0.141 &       &     &       &       &       &       &       \\ 
OV Boo	    &	DN? ZZ&	0.046258     &	0.0645&	 0.058&     &       &       &       &       &       \\ 
J1122-1110  &	DN AC &	0.04530	     &	0.088 &	 0.017&     &       &       &       &       &       \\ 
V396 Hya    &	NL AC &	0.04519	     & 	0.022 &	 0.017&	Y   &       &       &       &       &       \\ 
EI Psc 	    &	DN SU &	0.044566904  &	0.19  &	 0.13 &	Y   &	    &	YH  &	    &	Y   &       \\ 
1108+5728   &	DN SU &	0.03845	     &	      &	      &	Y   &       &       &       &       &       \\ 
GP Com	    &	NL AC &	0.032339     &	0.02  &	 0.011&	Y   &       &       &       &       &       \\ 
YZ LMi 	    &	DN AC &	0.019661     &	0.041 &	 0.035&     &       &       &       &       &       \\ 
V803 Cen    &	DN SU &	0.01848	     &	0.095 &	 0.074&     &       &       &       &       &       \\ 
KL Dra	    &	DN AC &	0.017382     &	0.075 &	 0.057&     &       &       &       &       &       \\ 
CR Boo	    &	DN AC &	0.017029     &	0.095 &	 0.064&     &       &       &       &       &       \\ 
HP Lib	    &	NL AC &	0.012763     &	0.12  &	 0.068&     &       &       &       &       &       \\ 
AM CVn	    &	NL AC &	0.011907     &	0.18  &	 0.13 &     &       &       &       &       &       \\ 
HM Cnc	    &	NL AC &	0.003721     &	0.50  &	 0.27 &     &       &       &       &       &       \\ 
\tableline 
\end{tabular} 
\\
\justify
The CV type, period, mass ratio, and secondary are from \citet{rit03}'s catalogue, which was last updated on Dec 31, 2015. Orbital periods for V603 Cas, J1544+2553, RW Sex, UU Aql, V380 Oph, QZ Ser, SU UMa, QZ Vir and EI Psc are from Thoretensen et al. (2017), for TW Vir from \citet{dai17}, for HU Aqr from \citet{sch18} and for AR UMa from \citet{bai16}. The latter also found that an AlI absorption line doublet in the visual varied. Orbital periods and other parameters for V1017 Sgr, QZ Aur, BT Mon, 
HR Del, DQ Her and RR Pic are updated by \citet{sch20}, for KR Aur by 
Rodriquez-Gil et al.(2020), for V1838 Aql by \citet{ech19}, for RW Tri by \citet{sub20}, for J0350+3232 by \citet{mas19}, for QZ Lib by \citet{pal18} and for OY Ara and V728 Sco by \citet{fue20}. The $q$ value and secondary mass for V396 Hya (CE 315) are from \citet{rui01}. A \enquote{Y} indicates an abundance anomaly was observed while a \enquote{N} means it was looked for but not found. \enquote{V} stands for a variable feature and a \enquote{YH} designates enhanced abundances of odd-numbered nuclear species above oxygen. 
\citet{har18} observations support a $q$ = 0.96 for DX And which we used instead of the listed 0.66.  The $q$ value for EX Hya is calculated
 from new observations of K1 \citep{ech16}. Using constraints from not having any superhumps on the upper limit of $q$ = 0.36 for DW Uma, \citet{pat05}
 give $q$ = 0.28$\pm$.04, which we use. The $q$ = 0.10 for MR Ser from Shahbaz \& Wood(1996) and the $q$ = 0.017 for J1122-1110 from Breedt et al.(2012) are included.
 Although VZ Scl has a Ritter-Kolb Catalogue listed $m_2$ of 1.4$M_{\odot}$ from Warner \& Thackeray (1975), a later paper (O'Donoghue et al. 1987)
 finds $m_2$ = 0.32$M_{\odot}$ which we used.  For V1223 Sgr, Penning(1985) and Watts et al.(1985) both using a main-sequence mass-radius relationship
 found $m_2$ = 0.40$M_{\odot}$.  Beuermann et al. (2004), noting that the secondary was bloated, commented on the effects of using a $m_2$ of 
 0.25$M_{\odot}$.  \citet{rit03} adopted an $m_2$ of $0.33 \pm 0.08M_{\odot}$ to cover this mass range.  For our analysis we adopted $m_2$ of 0.40$M_{\odot}$. We included QS Vir (EC13471-1258), since it is most likely a pre-CV just about to become a CV \citep{par16,odo03}.
 
\end{center}
\end{table} 

\end{document}